\documentclass[times,twocolumn,final]{elsarticle}

\usepackage{medima}
\usepackage{framed,multirow}

\usepackage{amssymb}
\usepackage{latexsym}

\usepackage{url}
\usepackage[table,xcdraw]{xcolor}
\colorlet{purple}{black}
\colorlet{yellow}{black}
\colorlet{blue}{black}
\usepackage{hyperref}

\usepackage{amsmath}
\usepackage{bbm}
\usepackage{upgreek}
\usepackage{adjustbox}
\usepackage{makecell}

\usepackage[switch]{lineno}

\definecolor{newcolor}{rgb}{.8,.349,.1}

\journal{Medical Image Analysis}

\begin{document}

\verso{Hong Liu \textit{et~al.}}

\begin{frontmatter}

\title{Simultaneous Alignment and Surface Regression Using Hybrid 2D-3D Networks for 3D Coherent Layer Segmentation of Retinal OCT Images with Full and Sparse Annotations}

\author[1,2,3]{Hong \snm{Liu}\fnref{fn1}}
\author[3]{Dong \snm{Wei}\fnref{fn1}}
\fntext[fn1]{Hong Liu and Dong Wei contributed equally. Hong Liu contributed to this work during an internship at Tencent.}
\author[3]{Donghuan \snm{Lu}}
\author[4]{Xiaoying \snm{Tang}}
\author[1]{Liansheng \snm{Wang}\corref{cor1}}
\cortext[cor1]{Corresponding author: lswang@xmu.edu.cn}
\author[3]{Yefeng \snm{Zheng}\vspace{-.5mm}}

\address[1]{School of Informatics, Xiamen University, Xiamen 361005, China}
\address[2]{National Institute for Data Science in Health and Medicine, Xiamen University, Xiamen 361005, China}
\address[3]{Jarvis Research Center, Tencent YouTu Lab, Shenzhen 518075, China}
\address[4]{Department of Electronic and Electrical Engineering, Southern University of Science and Technology, Shenzhen 518055, China\vspace{-.5mm}}

\received{1 May 2013}
\finalform{10 May 2013}
\accepted{13 May 2013}
\availableonline{15 May 2013}
\communicated{S. Sarkar}

\begin{abstract}
Layer segmentation is important to quantitative analysis of retinal optical coherence tomography (OCT).
Recently, deep learning based methods have been developed to automate this task and yield remarkable performance.
However, due to the large spatial gap and potential mismatch between the B-scans of an OCT volume, all of them were based on 2D segmentation of individual B-scans, which may lose the continuity and diagnostic information of the retinal layers in 3D space.
Besides, most of these methods required dense annotation of the OCT volumes, which is labor-intensive and expertise-demanding.
This work presents a novel framework based on hybrid 2D-3D convolutional neural networks (CNNs) to obtain continuous 3D retinal layer surfaces from OCT volumes, which works well with both full and sparse annotations. 
The 2D features of individual B-scans are extracted by an encoder consisting of 2D convolutions. These 2D features are then used to produce the alignment {\color{purple}displacement vectors} and layer segmentation by two 3D decoders coupled via a spatial transformer module.
Two 
losses are proposed to utilize the retinal layers’ natural property of being smooth for B-scan alignment and layer segmentation, respectively, and are the key to the semi-supervised learning with sparse annotation.
The entire framework is trained end-to-end. To the best of our knowledge, this is the first work that attempts 3D retinal layer segmentation in volumetric OCT images based on CNNs. 
Experiments on a synthetic dataset and {\color{blue}three} public clinical datasets show that our framework can effectively align the B-scans for potential motion correction, 
and achieves superior performance to state-of-the-art 2D deep learning methods in terms of both layer segmentation accuracy and cross-B-scan 3D continuity in both fully and semi-supervised settings, 
thus offering more clinical values than previous works.
\end{abstract}


\end{frontmatter}


\section{Introduction}\vspace{-1mm}
Optical coherence tomography (OCT)---a non-invasive imaging technique based on the principle of low-coherence interferometry---can acquire 3D cross-section images of human tissue at micron resolutions \citep{huang1991optical}.
Due to its micron-level axial resolution, non-invasiveness, and fast speed, OCT is commonly used in eye clinics for diagnosis and management of retinal diseases \citep{abramoff2010retinal}.
Notably, OCT provides a unique capability to directly visualize the stratified structure of the retina of cell layers, whose statuses are biomarkers of presence, severity, and prognosis for a variety of retinal and neurodegenerative diseases, including age-related macular degeneration \citep{keane2009evaluation}, diabetic retinopathy \citep{bavinger2016effects}, glaucoma \citep{kansal2018optical}, Alzheimer’s disease \citep{knoll2016retinal}, and multiple sclerosis \citep{saidha2011primary}.
Usually, layer segmentation is the first step in quantitative analysis of retinal OCT images, yet can be considerably labor-intensive, time-consuming, and subjective when done manually.
Therefore, computerized tools for automated, prompt, objective, and accurate retinal layer segmentation in OCT images are highly desired.

Automated layer segmentation in retinal OCT images has long been explored.
Earlier explorations \citep{garvin2009automated,yazdanpanah2009intra,lang2013retinal} relied on empirical rules and/or hand-crafted features, which may be difficult to generalize.
Recently, researchers started to implement deep convolutional neural networks (CNNs) for retinal layer segmentation in OCT images and achieved superior performance to classical methods \citep{he2019fully, xie2022globally}.
However, most previous methods (both classical and CNNs) segmented each OCT slice (called a B-scan) separately given the relatively big inter-B-scan distance, 
despite the fact that a modern OCT sequence actually consists of many B-scans covering a volumetric area of the eye \citep{drexler2008state}.
Correspondingly, these methods failed to utilize the anatomical prior that the retinal layers are generally smooth surfaces (instead of independent curves in each B-scan), and may be subject to discontinuity in the segmented layers between adjacent B-scans, potentially affecting volumetric analysis following layer segmentation.
Although some works \citep{antony2013combined,carass2014multiple,chen2018intraretinal,garvin2009automated,lang2013retinal,novosel2017joint} attempted 3D OCT segmentation,
all of them belong to the classical methods that yielded inferior performance to the CNN-based ones, and overlooked the misalignment artifact of the B-scans in an OCT volume.

The inter-B-scan misalignment happens unavoidably mainly because of the involuntary eye movements during acquisition time \citep{sanchez2019review}.\footnote{Intra-B-scan misalignment often can be ignored given the fast A-scan acquisition rate of modern spectral domain OCT systems  \citep{mcnabb2012distributed}.
}
{\color{blue}The motion artefacts may adversely affect qualitative interpretation and quantitative analysis of the images.
For example, they may be mistaken for pathologies distorting the retinal pigment epithelium, and affect clinical decisions and the tracking of fine-grained features such as cysts or vessels between B-scans \citep{montuoro2014motion}.
In addition, the artefacts may distort the underlying 3D structures.
Such distortion may pose challenges in some applications, e.g., multi-modality registration where an OCT en face image and a color fundus image need to be aligned \citep{cheng2016motion}, 3D reconstruction and analysis of layer surface/thickness maps \citep{hood2014improving,janez2019spatial}, and 3D OCT segmentation where raw 3D operations may be invalidated in the presence of misalignment between B-scans.}

Besides the motion artefact, {\color{blue}another obvious obstacle to developing a CNN-based method for 3D OCT segmentation is the apparent anisotropy in resolution \citep{shah2018multiple}.}
For example, the physical resolutions of one of the datasets employed in this work are 3.24 $\upmu$m (within A-scan, which is a column in a B-scan image),  6.7 $\upmu$m (cross-A-scan), and 67 $\upmu$m (cross-B-scan).
Given the {\color{yellow}void of any existing CNN-based 3D OCT segmentation method}, it is therefore not strange that the anisotropy problem has not been considered in such a context before.

In this work, we propose a novel CNN-based 2D-3D hybrid framework for simultaneous B-scan alignment and 3D surface regression for coherent retinal layer segmentation across-B-scan in OCT images.
This framework consists of a shared 2D encoder followed by two 3D decoders (the alignment branch and the segmentation branch), and a spatial transformer module \citep[STM;][]{balakrishnan2019voxelmorph} inserted to the shortcuts \citep{ronneberger2019u} between the encoder and the segmentation branch.
Given a B-scan volume as input, we employ per B-scan 2D operations for the encoder for two reasons.
First, {\color{blue}as suggested by previous studies \citep{zhang2019light,wang2020conquering},
intra-slice feature extraction followed by inter-slice (2.5D or 3D) aggregation is an effective strategy against anisotropic resolution, thus we propose a similar 2D-3D hybrid structure for the anisotropic OCT data}.
Second, the B-scans in the input volume are subject to misalignment, thus 3D operations across-B-scan prior to proper realignment may be invalid.
Following the encoder, the alignment branch employs 3D operations to aggregate features across-B-scan to align them properly.
Then, the resulting {\color{purple}displacement vectors} are employed
to align the 2D features at different scales and compose well-aligned 3D features by the STM.
These 3D features are passed to the segmentation branch for 3D surface regression.
Noteworthily, the alignment only ensures validity of subsequent 3D operations, but not necessarily the cross-B-scan coherence of the regressed layer surfaces.
Hence, we further impose a gradient-based, 3D regulative loss \citep{wei2018three} on the regressed surfaces to encourage surface smoothness, which is an intrinsic property of many biological layers;
we refer to this loss as the \emph{global coherence loss}.
While it is straightforward to implement the global coherence loss within our surface regression framework and comes for free (no manual annotation is needed), it proves effective in our experiments.
The entire framework is trained end-to-end.

Last but not least, we are delighted to discover that our proposed 2D-3D framework can naturally handle a practical scenario of semi-supervised learning for OCT layer segmentation, where only a subset of B-scans in each OCT volume is manually annotated (i.e., sparse annotation).
Owing to the introduction of the global coherence loss, layers segmented in non-annotated B-scans can be optimized according to their coherence with these layers in a neighborhood of B-scans---no matter annotated or not.
In such scenario, the segmentation and alignment branches are tangled even more closely than in the fully supervised setting, mutually benefiting each other.
Considering the labor-intensive and time-consuming nature of the manual layer annotation, effective semi-supervised learning is especially valuable.
It should be noted that although few other works also attempted semi-supervised OCT layer segmentation \citep{liu2018semi,sedai2019uncertainty}, they all treated the B-scans as independent 2D images and relied on the notion of uncertainty/confidence.
So far as we are aware of, our work is the first that bases semi-supervised OCT layer segmentation on 3D coherence of the layers.

In summary, our contributions are as follows:
\begin{itemize}
  \item First, we propose a new framework for simultaneous B-scan alignment and 3D layer segmentation of retinal OCT images.
      This framework features a hybrid 2D-3D structure comprising a shared 2D encoder, a 3D alignment branch, a 3D surface regression branch, and an STM to allow for simultaneous alignment and 3D segmentation of anisotropic OCT data.
  \item Second, we further incorporate a conceptually straightforward and easy-to-implement regulating loss, the global coherence loss, to encourage the regressed layer surfaces to be coherent---not only within but also across-B-scan.
  {\color{purple}Jointly, the first two contributions enable our framework to produce \textit{more coherent layer surfaces in 3D} than existing state-of-the-art \citep{he2019fully,he2021structured,xie2022globally,xie2022deep}, as validated by our experiments.
  This advantage makes our framework preferred in applications where 3D fidelity of the segmented structures is crucial, e.g., 3D reconstruction and analysis of layer surface/thickness maps.}
  \item Third, we extend the framework for semi-supervised learning where only a subset of B-scans in each OCT volume is annotated.
  Thanks to our novel design of coupled B-scan alignment and 3D layer segmentation, and the global coherence loss, the extension is straightforward yet remarkably effective.
  {\color{purple}Experiments show that the performance advantages of our framework over other methods become more prominent with decreasing number of B-scans annotated and demonstrate its practical usability given sparse annotations.
  This capability of semi-supervised learning is valuable considering the effort and difficulty of manual labeling.}
\end{itemize}
We conduct thorough experiments on {\color{blue}three} public OCT datasets as well as synthetic data, to evaluate effectiveness of the proposed framework, validate its design, and demonstrate its superiority toward existing methods in terms of both B-scan alignment and fully/semi-supervised segmentation.

This work is a comprehensive extension to our proof-of-concept exploration \citep{liu2021simultaneous} in three main aspects, i.e., we (1) extend the framework to support semi-supervised learning, (2) additionally use synthetic data to quantify the B-scan alignment performance, and (3) employ {\color{blue}two more public datasets} to evaluate the generalization of the proposed framework.

\section{Related work}

\subsection{Retinal OCT segmentation}
Earlier attempts at automated retinal layer segmentation in OCT images included graph based \citep{antony2013combined,garvin2009automated,lang2013retinal}, 
contour modeling \citep{carass2014multiple,novosel2017joint,yazdanpanah2009intra}, and machine learning \citep{antony2013combined,lang2013retinal} methods.
{\color{blue}For example, the graph theory and dynamic programming framework \citep{chiu2010automatic}, an inferential classic approach, modelled the layer segmentation problem as finding the shortest path in a graph representing the OCT image.}
Although greatly advanced the field,
most of these classical methods relied on empirical rules and/or hand-crafted features which may be difficult to generalize.
Motivated by the success of deep convolutional neural networks (CNNs) in a wide variety of medical image analysis tasks~\citep{ker2017deep,litjens2017survey,shen2017deep}, researchers also implemented CNNs for retina OCT segmentation and achieved superior performance to classical methods,
mainly attributed to the data-driven automatic extraction of task-appropriate features.
\cite{fang2017automatic} and \cite{kugelman2018automatic} conducted graph search on CNN-based probability maps.
\cite{liu2018automated} trained a structured random forest classifier on integrated deep CNN and hand-crafted features.
These works relied on patch-based classification of (the local neighborhood of) each pixel of interest.
More recently, fully convolutional networks (FCNs) \citep{long2015fully} were applied to retina OCT  segmentation, achieving great improvement in both efficiency and accuracy \citep{roy2017relaynet,shah2018multiple,he2019fully,he2021structured}.
\cite{li2021multi} employed graph based representations \citep{atif2007generic} to assist FCNs in exploiting anatomical prior knowledge and performing spatial reasoning.
{\color{blue}\citet{xie2022globally} proposed to explicitly enforce mutual surface interaction constraints with a graph model and realize simultaneous total surface cost minimization and surface order constraints with a primal-dual interior-point method (IPM).
\citet{xie2022deep} proposed to integrate a constrained differentiable dynamic programming (DDP) module in end-to-end training to enforce surface smoothness.}
Our method also belongs to the FCN genre.
However, distinct from all the FCN-based methods above which segmented individual B-scans as independent 2D images separately, our method segments all B-scans in the same OCT volume together in 3D, after aligning them properly within the same framework.

\subsection{OCT motion correction}

Correction of involuntary eye motion in retinal OCT can be accomplished with either a hardware or software solution \citep{baghaie2017involuntary}.
%
Hardware correction can be performed in an online or off-line manner, and the correction effects are promising \citep{ferguson2004tracking,vienola2012real,kocaoglu2014adaptive}.
%
However, it requires special hardware not broadly available in current clinic practice, and cannot be applied to legacy data.
%

Alternatively, software-based postprocessing provides an economic solution.
%
%
\cite{capps2011correction}, \cite{xu2009shape}, and \cite{ricco2009correcting} used scanning laser ophthalmoscopy 
or color fundus images acquired along with the OCT scans for correcting the transverse motion, and \cite{kraus2014quantitative,antony2011automated} used additional orthogonal scans to help reconstruct true curvature of the retina.
Needing extra scans as alignment reference, however, these methods added complexity to the imaging process and still had a limited applicability to a large amount of legacy data.
%
\cite{montuoro2014motion} assumed local symmetry for the shape of the retina and eliminated the need for any auxiliary scan, yet the assumption can be violated in pathological areas and in the proximity of the fovea.
More recently, segmentation---e.g., of background, retinal layers, or vessels---guided motion correction was proposed \citep{montuoro2014motion,fu2016eye,lezama2016segmentation}.
%
However, the segmentations in the previous works were often coarse, independent of the motion correction, and 2D in nature (as 3D segmentation cannot be done with validity prior to proper B-scan alignment).
%
In contrast, we couple precise 3D layer segmentation with motion correction for effective mutual performance boosting.

\subsection{Hybrid 2D-3D networks}
To leverage the strengths of both 2D and 3D networks for volumetric image analysis, i.e., parameter (and computation) efficiency and inter-slice correlation, respectively, hybrid 2D-3D networks were proposed \citep{li2018h,tran2018closer,wang2019automatic,xie2018rethinking,zhang2019light}. 
%
Besides, the hybrid architecture allowed the use of existing pretrained 2D network parameters for effective transfer learning \citep{li2018h,wang2020conquering}.
Several studies showed that hybrid 2D-3D networks were also suitable for volume segmentation of anisotropic resolutions \citep{li2018h,wang2019automatic,zhang2019light}, where individual slices were first processed by 2D operations to yield features appropriate for subsequent 3D operations.
%
However, none of them considered inter-slice motion artifacts and thus could not be directly applied to 3D segmentation of OCT volumes.
In this work, besides employing a hybrid 2D-3D architecture to deal with the data anisotropy, we additionally rely on the same architecture to correct the inter-B-scan misalignment simultaneously.

\begin{figure*}[!t]
\centering
\includegraphics[width=.95\textwidth,trim=0 0 0 0,clip]{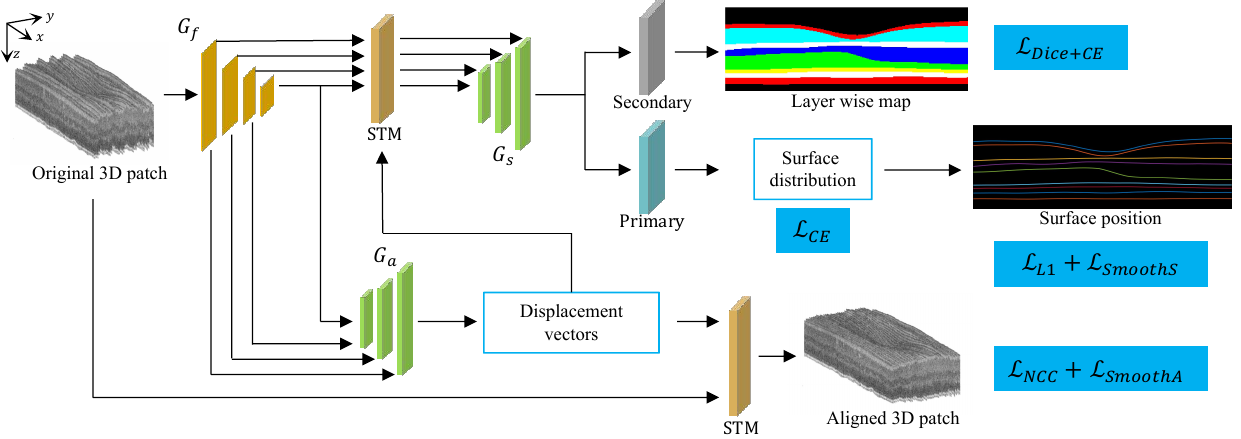}
\caption{\color{purple}Overview of the proposed framework.} \label{fig:overview}
\end{figure*}

\subsection{Semi-supervised medical image segmentation with sparse annotation}
To ease the heavy burden of manual volumetric segmentation labeling, many methods have been proposed for semi-supervised medical image segmentation with sparse annotation.
Some of these methods employed non-rigid registration for label prorogation \citep{bai2018recurrent}, conducted self-training for pseudo label generation \citep{zheng2020cartilage}, or combined both \citep{bitarafan20203d}.
A central idea of these semi-supervised methods was the uncertainty estimation, based on which the pseudo labels were filtered or weighted.
%
%
Several methods were also proposed for semi-supervised segmentation of retinal layers in OCT images.
\cite{sedai2019uncertainty} proposed uncertainty guided semi-supervised learning based on a teacher-student approach, whereas \cite{liu2018semi} proposed to estimate the uncertainty with a discriminator based on adversarial learning.
Unlike these works relying on pseudo label generation and uncertainty estimation, our method makes use of a physical property of the segmentation target (i.e., 3D coherence of the retinal layers) and also benefits from the entangled 3D surface regression and B-scan alignment.
Thanks to the unique problem formulation, it can accommodate both fully and sparsely annotated segmentation with the same framework
and are competent in both cases.

\section{Problem formulation}

Let $\Omega\subset\mathbb{R}^3$, then a 3D OCT volume can be written as a real-valued function $V(x,y,z):\Omega\rightarrow\mathbb{R}$, where the $x$ and $z$ axes are the row and column directions of a B-scan image, and $y$ axis is orthogonal to the B-scan image (see the illustration on the far left of Fig. \ref{fig:overview}).\footnote{Note the definition of axes is different from what is commonly used in axial CT and MRI volumes, where the $z$ axis is orthogonal to the imaging planes.}
Alternatively, $V$ can be considered as an ordered collection of all its B-scans: $V=\{I_b\}$, where $I_b:\Phi\rightarrow\mathbb{R}$ is the $b$\textsuperscript{th} B-scan image, $\Phi\subset\mathbb{R}^2$, $b\in[1,N_B]$, and $N_B$ is the number of B-scans.
Then, a retinal layer surface can be expressed by $S(b,a)=r_{b,a}:\Psi\rightarrow\mathbb{R}$, where $\Psi\subset\mathbb{R}^2$, $a\in[1,N_A]$, $N_A$ is the number of A-scans, and $r_{b,a}$ is the row index indicating the surface location in the $a$\textsuperscript{th} A-scan of the $b$\textsuperscript{th} B-scan.
That is, the surface intersects with each A-scan exactly once.
Due to the image acquisition process wherein each B-scan is acquired separately without a guaranteed global alignment and the inevitable eye movement, consecutive B-scans in an OCT volume may be subject to misalignment \citep{cheng2016motion}.
%
The misalignments along the $x$ and $z$ axes are also known as the transverse and axial motion artifacts, respectively.
Therefore, the goal of this work is to locate a set of retinal layer surfaces of interest $\{S\}$---preferably being smooth---for accurate segmentation of the layers, while at the same time re-aligning the set of B-scans $\{I_b\}$ in $V$.
{\color{yellow}It is worth mentioning that as the transverse motion only occurs on several B-scans (meaning most of the B-scans have no $x$ movement) \citep{fu2016eye}, we only correct the axial motion within the proposed simultaneous alignment and surface regression framework, but (optionally) correct the transverse with simple post-processing.}

\section{Method}

\subsection{Overview}
The overview of our framework is shown in Fig. \ref{fig:overview}.
The framework comprises three major components: a contracting path $\mathnormal{G_f}$ (the shared encoder) consisting of 2D CNN layers and two expansive paths consisting of 3D CNN layers $\mathnormal{G_a}$ (the alignment branch) and $\mathnormal{G_s}$ (the segmentation branch), and a functional module: the spatial transformer module (STM).
First, 2D features of separate B-scans in an OCT volume are extracted by $\mathnormal{G_f}$.
These features are then used to generate B-scan alignment displacement by $\mathnormal{G_a}$, which is used by the STM in turn to align the 2D features.
After that, the well-aligned features are fed to $G_s$ to yield final segmentation.
Each of $G_a$ and $G_s$ forms a hybrid 2D-3D residual U-Net \citep{ronneberger2019u} with $G_f$.
The difference with the general U-Net is that our U-Net consists of both 2D and 3D CNN networks. 
The entire framework is trained end-to-end.
{\color{yellow}As $G_f$ is implemented as a routine 2D CNN feature extractor, 
below we focus on describing our novel $G_a$, $G_s$, and STM.}

\subsection{B-Scan alignment branch}
Although it is possible to add an alignment step while preprocessing, a comprehensive framework that couples the B-scan alignment and layer segmentation would mutually benefit each other (supported by our experimental results), besides being more integrated.
To this end, we introduce a B-scan alignment branch consisting of an expansive path into our framework, which takes 2D features extracted from a set of B-scans by $G_f$ and outputs a displacement vector $\triangle\boldsymbol{d}=[d_1, \ldots, d_{N_B}]$, with each element $d_b$ indicating the displacement for a B-scan in the $z$ direction.
As smoothness is one of the intrinsic properties of the retinal layers, if the B-scans are aligned properly, ground truth surface positions of the same layer should be close at nearby locations of adjacent B-scans.
To model this prior,
we propose a supervised loss function to help with the alignment:
\begin{linenomath*}
\begin{equation}\label{eq:smooth_a}
    \mathcal{L}_\mathrm{SmoothA}={\sum}_{b=1}^{N_B-1}{\sum}_{a=1}^{N_A}
    \big((r^g_{b,a}-d_{b}) - (r^g_{b+1,a}-d_{b+1})\big)^2,
\end{equation}
\end{linenomath*}
where $r^g$ is the ground truth surface location.

Meanwhile, we also use the local normalized cross-correlation (NCC) \citep{balakrishnan2019voxelmorph} of adjacent B-scans as the unsupervised optimization objective of $\mathnormal{G_a}$:
\begin{linenomath*}
\begin{equation}\color{yellow}\small
    \mathcal{L}_\mathrm{NCC}=\sum_{b=1}^{N_B-1} \sum_{\boldsymbol{p}\in{\Phi}} \frac{\Big[\sum_{\boldsymbol{p}_k}\big(\hat{I}_{b}(\boldsymbol{p}_k)-\bar{I}_{b}(\boldsymbol{p})\big)\big(\hat{I}_{b+1}(\boldsymbol{p}_k)-\bar{I}_{b+1}(\boldsymbol{p})\big)\Big]^2}
    {\big[\sum_{\boldsymbol{p}_k}(\hat{I}_{b}(\boldsymbol{p}_k)-\bar{I}_{b}(\boldsymbol{p}))^2\big]\big[\sum_{\boldsymbol{p}_k}(\hat{I}_{b+1}(\boldsymbol{p}_k)-\bar{I}_{b+1}(\boldsymbol{p}))^2\big]},
\end{equation}
\end{linenomath*}
where $\boldsymbol{p}$ iterates over all pixels in the image space $\Phi$, $\hat{I}_b$ is the $b$\textsuperscript{th} B-scan image displaced according to the corresponding $d_b$ (described in the following section), and $\bar{I}$ denotes images with local mean intensities subtracted out: ${\bar{I}(\boldsymbol{p})=\hat{I}(\boldsymbol{p})-\frac{1}{n^2}\sum_{\boldsymbol{p}_k}\hat{I}(\boldsymbol{p}_k)}$, where $\boldsymbol{p}_k$ iterates over an $n\times n$ region around $\boldsymbol{p}$.
We follow \citep{balakrishnan2019voxelmorph} to set $n=9$.
The final optimization objective of the alignment branch is:
\begin{linenomath*}
\begin{equation}
\mathcal{L}_\mathrm{Align}=\mathcal{L}_\mathrm{NCC} + \mathcal{L}_\mathrm{SmoothA}.
\end{equation}
\end{linenomath*}

\subsection{Spatial transformer module}\label{sec:STM}
Besides being used to align the input B-scan images, the displacement vector $\triangle\boldsymbol{d}$ output by the alignment branch $G_a$ is also used to align the 2D features extracted by $G_f$, such that subsequent 3D operations of the segmentation branch $G_s$ are valid.
To do so, we propose to add a spatial transformer module (STM) \citep{balakrishnan2019voxelmorph} to the shortcuts between $G_f$ and $G_s$.
It is worth noting that the STM adaptively rescales $\triangle\boldsymbol{d}$ to suit the size of the features at different scales.
Without loss of generality, we use the input B-scan images for explanation.
Specifically, for each pixel $\boldsymbol{p}=(p_x, p_z)$ in the relocated B-scan image $\hat{I}_b$, we compute a (sub-)pixel location {\color{yellow}$\boldsymbol{p}'=\boldsymbol{p}+(0,d_b)$} (recall that we only consider axial motion in the networks) in the original image $I_b$.
Then, we linearly interpolate the values at neighboring pixels of $\boldsymbol{p}'$ as the value for $\hat{I}_b(\boldsymbol{p})$:
\begin{linenomath*}
\begin{equation}
    \hat{I}_b(\boldsymbol{p}) = {\sum}_{\boldsymbol{q}\in\mathcal{Z(\boldsymbol{p}')}}
    I_b(\boldsymbol{q})\big(1-|\boldsymbol{p}'_x-\boldsymbol{q}_x|\big)\big(1-|\boldsymbol{p}'_z-\boldsymbol{q}_z|\big),
\end{equation}
\end{linenomath*}
where $\mathcal{Z}(\boldsymbol{p}')$ are the pixel neighbors of $\boldsymbol{p}'$.
The STM 
allows back prorogation during optimization \citep{balakrishnan2019voxelmorph}.
The application of $\triangle\boldsymbol{d}$ to the 2D features is mostly the same, except for that $\triangle\boldsymbol{d}$ is rescaled to suit the downsampling factors of the features at different scales.
In this way, we couple the B-scan alignment and retinal layer segmentation in our framework for an integrative end-to-end training, which not only simplifies the entire pipeline but also boosts the segmentation performance as validated by our experiments.

\subsection{Layer segmentation branch}
Our layer segmentation branch substantially extends the fully convolutional boundary regression (FCBR) framework proposed by \cite{he2019fully}.
Above all, we replace the purely 2D FCBR framework by a hybrid 2D-3D framework, to perform 3D surface regression in an OCT volume instead of independent 2D boundary regression in individual B-scans.
On top of that, we propose a global smoothness guarantee loss to encourage coherent surfaces both within and across-B-scan, whereas FCBR only enforces intra-B-scan smoothness.
Third, our segmentation branch is coupled with the B-scan alignment branch, which boosts the performance of each other.

The segmentation branch has two heads sharing the same decoder: the primary head outputs the surface position distribution for each A-scan, and the secondary head outputs pixel-wise semantic labels.
The secondary head is used only to provide an additional task 
for training the network, especially considering its pixel-wise dense supervision.
Eventually the output of the secondary head is ignored during testing.
We follow He \emph{et al.} to use a combined Dice and cross entropy loss \citep{roy2017relaynet} $\mathcal{L}_\mathrm{Dice+CE}$ for training the secondary head, and refer interested reader to \citep{he2019fully} for more details.

\paragraph{Surface distribution head}
This primary head generates an independent surface position distribution $q_{b,a}(r|V;\theta)$ for each A-scan, where $\theta$ denotes the network parameters, and a higher value indicates a higher possibility that the surface is on the $r$\textsuperscript{th} row.
Like in \citep{he2019fully}, a cross entropy loss is used to train the primary head:
\begin{linenomath*}
\begin{equation}
    \mathcal{L}_\mathrm{CE}=-{\sum}_{b=1}^{N_B}{\sum}_{a=1}^{N_A}{\sum}_{r=1}^{R}
    \mathbbm{1}(r^g_{b,a}=r)\log{ q_{b,a}(r_{b,a}^g|V,\theta)},
\end{equation}
\end{linenomath*}
where $R$ is the number of rows of an A scan, and $\mathbbm{1}(x)$ is the indicator function where $\mathbbm{1}(x)=1$ if $x$ is evaluated to be true and zero otherwise.
Further, a smooth L1 loss is adopted to directly guide the predicted surface location $\hat{r}$ to be the ground truth:
\begin{linenomath*}
\begin{equation}
    \mathcal{L}_\mathrm{L1} = {\sum}_{b=1}^{N_B}{\sum}_{a=1}^{N_A} 0.5t_{b,a}^2\mathbbm{1}(|t_{b,a}|<1)+(|t_{b,a}|-0.5)\mathbbm{1}(|t_{b,a}|\geq1),
\end{equation}
\end{linenomath*}
where $t_{b,a}=\hat{r}_{b,a}-r_{b,a}^g$, and $\hat{r}_{b,a}$ is obtained via the soft-argmax
$
    \hat{r}_{b,a}={\sum}_{r=1}^R rq_{b,a}(r|V,\theta).
$



\paragraph{Global coherence loss}
Previous studies have demonstrated the effectiveness of modeling prior knowledge that reflects anatomical properties such as the structural smoothness \citep{wei2018three} in medical image segmentation.
Following this line, we also employ a global smoothness loss to encourage the detected retinal surface $\hat{S}$ to be coherent both within and across-B-scan based on its gradients:
\begin{linenomath*}
\begin{equation}\color{purple}
    \mathcal{L}_\mathrm{SmoothS} = {\sum}_{b=1}^{N_B}{\sum}_{a=1}^{N_A}\big\|\triangledown \hat{S}(b,a)\big\|^2.
\label{Grad}
\end{equation}
\end{linenomath*}
Finally, the overall optimization objective of the segmentation branch is 
\begin{linenomath*}
\begin{equation}\label{eq:L_seg}
\mathcal{L}_\mathrm{Seg}=\mathcal{L}_\mathrm{Dice+CE}+\mathcal{L}_\mathrm{CE}+\mathcal{L}_\mathrm{L1}+\lambda \mathcal{L}_\mathrm{SmoothS},
\end{equation}
\end{linenomath*}
where $\lambda$ is a hyperparameter controlling the influence of the global coherence loss.





\subsection{Semi-supervised learning with sparse annotations}\label{sec:method:semi}

Next, we extend our framework for semi-supervised learning with sparse annotations.
To reduce the human efforts for labeling whole OCT volumes,
we propose to further leverage the smoothness property of the retinal layers, so that the segmentation model can be effectively trained with only a fraction of the B-scans annotated in each given OCT volume.
Thanks to our unique problem formulation of coupled 3D surface regression and B-scan alignment, the adaptation for the semi-supervised setting is straightforward with only minor alterations to the loss functions.
For the B-scan alignment branch, we adapt the supervised loss $\mathcal{L}_\mathrm{SmoothA}$ (Eqn. (\ref{eq:smooth_a})) by using the surface locations predicted by $G_s$ for unannotated B-scans: 
\begin{linenomath*}
\begin{equation}\label{eq:smooth_a_semi}
    \mathcal{L}^\mathrm{Semi}_\mathrm{SmoothA}={\sum}_{b=1}^{N_B}{\sum}_{a=1}^{N_A}
    \big((r_{b,a}-d_{b}) - (r_{b+1,a}-d_{b+1})\big)^2,
\end{equation}
\end{linenomath*}
where $r_{b,a}=\hat{r}_{b,a}$ for unannotated B-scans and $r^g_{b,a}$ otherwise, while the unsupervised alignment loss $\mathcal{L}_\mathrm{NCC}$ remains unchanged.
To this end, $\mathcal{L}^\mathrm{Semi}_\mathrm{SmoothA}$ couples $\mathnormal{G_s}$ and $\mathnormal{G_a}$ in a more delicate way, where the segmentation and alignment results interactively influence each other.
As to the layer segmentation branch, we now calculate the supervised losses 
$\mathnormal{\mathcal{L}_\mathrm{Dice+CE}}$, $\mathnormal{\mathcal{L}_\mathrm{L1}}$ and $\mathnormal{\mathcal{L}_\mathrm{CE}}$ only on the annotated B-scans, with the regulating global coherence loss $\mathcal{L}_\mathrm{SmoothS}$ intact. 

It is worth mentioning that as $\mathcal{L}^\mathrm{Semi}_\mathrm{SmoothA}$ (Eqn. (\ref{eq:smooth_a_semi})) relies on the model-predicted surface location $\hat{r}_{b,a}$, we notice that bad quality of $\hat{r}_{b,a}$ would impede the training or even cause a collapse.
Therefore, we first warm up the model without $\mathcal{L}^\mathrm{Semi}_\mathrm{SmoothA}$ for five epochs, such that the predicted $\hat{r}_{b,a}$ is reasonable when adding $\mathcal{L}^\mathrm{Semi}_\mathrm{SmoothA}$ back afterwards.

\subsection{Transverse alignment by post-processing}
The transverse motions are much less frequent than the axial for the OCT B-scans \citep{fu2016eye}.
Notwithstanding, for completeness, we still propose a simple yet effective post-processing transverse alignment for optional use when necessary.
Specifically, we first average the A-scans of each B-scan image to turn the latter into a strip of mean intensity projections.
We then align adjacent B-scans by shifting them in the $x$ direction to minimize the mean squared error between their projections.
{\color{purple}Notably, as we already have the retinal layers segmented, we eliminate the interference of the background by computing the mean projection only in the retinal layers.}
The entire OCT volume is aligned by repeating the pairwise alignment.


\section{Experiments}

\subsection{Datasets and preprocessing}

The proposed framework is validated on {\color{blue}three} publicly available
SD-OCT datasets.
In addition, synthetic images are utilized for quantitative evaluation of the inter-B-scan motion correction.
Below we describe the public datasets first, and defer the description of the synthetic data to the corresponding experiments section.

\paragraph{A2A SD-OCT Study dataset} 
The Age-Related Eye Disease Study 2 (AREDS2) Ancillary SD-OCT (A2A SD-OCT) Study dataset \citep{farsiu2014quantitative} includes both normal {\color{black}(115)} and age-related macular degeneration (AMD) {\color{black}(269)} cases.
The images were acquired using the Bioptigen Tabletop SD-OCT system (Bioptigen, Inc., Research Triangle Park, NC).
The physical resolutions are 3.24 $\upmu$m (within A-scan), 6.7 $\upmu$m (cross-A-scan), and 67 $\upmu$m (cross-B-scan).
Since the manual annotations are only available for a region centered at the fovea, subvolumes of size $400\times41\times512$ ($N_A$, $N_B$, and $R$) voxels are extracted around the fovea.
We train the model on 263 subjects and test on the other 72 subjects ({\color{yellow}49} cases are eliminated from analysis as the competing alignment algorithm \citep{pnevmatikakis2017normcorre} fails to handle them), which are randomly split with the proportion of AMD cases unchanged.
The inner aspect of the inner limiting membrane (ILM), inner aspect of the retinal pigment epithelium drusen complex (IRPE), and outer aspect of Bruch's membrane (OBM) were manually traced.

\paragraph{JHH SD-OCT dataset} The Johns Hopkins Hospital (JHH) dataset \citep{he2019retinal} contains 14 healthy controls (HC) and 21 cases with multiple sclerosis (MS) which exhibit mild thinning of retinal layers. 
The data was acquired with a Spectralis OCT system  (Heidelberg Engineering, Heidelberg, Germany), which has its own motion correction, registration, and averaging algorithms during image acquisition.
Each of the 35 cases includes 49 B-scans of $1024\times496$ ($N_A\times R$) pixels in size.
The physical resolutions are 3.87 $\upmu$m (within A-scan), 5.8 $\upmu$m (cross-A-scan), 123.6 $\upmu$m (cross-B-scan).
Following the train/test split in \citep{he2019fully}, we use the last six HCs and last nine MS cases for training and the other 20 subjects for testing.
Nine surfaces were manually delineated in each B-scan, separating the following retinal layers: the retinal nerve fiber layer (RNFL); the ganglion cell layer (GCL) combined with the inner plexiform layer (IPL), denoted as GCIP; the inner nuclear layer (INL); the outer plexiform layer (OPL); the outer nuclear layer (ONL); the inner segment (IS); the outer segment (OS); and the retinal pigment epithelium (RPE). Surfaces between these layers are denoted by hyphenating their acronyms.
Three other named surfaces are: the inner limiting membrane (ILM);
the external limiting membrane (ELM); and Bruch’s membrane (BM).

{\color{blue}\paragraph{Duke DME dataset} The Duke Eye Center dataset \citep{Chiu:15} contains 10 diabetic macular edema (DME) patients (one OCT volume per patient), each with 61 B-scans of size $768\times496$ ($N_A\times R$) pixels.
The first five patients were rated as having severe macular edema with damaged retinal structures.
The data was acquired with a standard Spectralis (Heidelberg Engineering, Heidelberg, Germany) 61-line volume scan protocol.
The physical resolutions are 3.87 $\upmu$m (within A-scan), 10.94--11.98 $\upmu$m (cross-A-scan), and 118--128 $\upmu$m (cross-B-scan).
Eight retinal surfaces (the same ones as those delineated on the JHH dataset except for ELM) 
were manually delineated for 11 B-scans per patient.
We follow the 50\%:50\% train/test split in previous works \citep{Chiu:15,he2021structured,karri2016learning,rathke2017locally} to use the last five patients for training and the challenging first five patients for testing.
As our framework takes 3D volumes as input, we use the partially annotated OCT volumes (i.e., 11 of 61 B-scans annotated per volume) in this dataset as semi-supervised learning with sparse annotations (cf. Section \ref{sec:method:semi}).}

\paragraph{Preprocessing} An intensity gradient method \citep{lang2013retinal} is employed to flatten the B-Scan images to the estimated Bruch's membrane.
After that, B-scan images in the A2A, JHH, and DME datasets are cropped to $400\times320$, $1024\times128$ and {\color{blue}$768\times224$} pixels, respectively, to exclude background while ensuring inclusion of retinal tissue \citep{he2021structured}.
The preprocessing effectively reduces memory consumption for model training.

\subsection{Evaluation metrics}
For B-scan alignment, we adopt the mean absolute distance (MAD) of the same surface and the NCC between two adjacent B-scans
for quantitative evaluation on the A2A dataset.
In addition, on the synthetic dataset we directly calculate the mean absolute difference 
between the estimated and ground truth motions (note that the motions are simulated in this setting thus the ground truth is available) for evaluation.
For retinal layer segmentation, the MAD {\color{blue}and 95\textsuperscript{th} percentile
of the Hausdorff distance (HD95)} between predicted and ground truth surface positions 
are used.
To quantify the cross-B-scan continuity of the segmented surfaces, inspired by \cite{he2021structured}, we compute the surface distances between adjacent B-Scans as the statistics of smoothness and plot the histogram for visual analysis.
{\color{purple}We compute the mean metrics and standard deviations (std.) per volume, per the volumetric nature of our proposed framework.
Especially for MAD, all A-scan-wise differences of an OCT volume are first averaged to yield a volume-wise metric.
Then, a mean metric and std. are estimated from the volume-wise metrics of all test volumes.}
{\color{blue} Note that for the DME dataset, we follow \citep{he2019fully,he2021structured} to ignore the positions where \citep{Chiu:15}'s result or the manual delineation are missing for evaluation.}

\subsection{Implementation}
The PyTorch framework (1.4.0) is used for all experiments.
For the network design, we mainly follow the architecture proposed in Model Genesis \citep{zhou2019models} with necessary adaptations:
(i) the feature extractor $G_f$ is constructed by replacing 3D operations of the Model Genesis's encoder with 2D counterparts,
(ii) the alignment and segmentation branches $G_a$ and $G_s$ are mostly the same as the decoder in Model Genesis, except that we keep the dimension corresponding to the number of B-scans unchanged throughout, and
%
(iii) we halve the number of channels in each CNN block to reduce the number of network parameters.
All networks are trained from scratch.
Due to the GPU memory constraint, the strategy of patch-wise training is employed: the OCT volumes are cut by planes perpendicular to the B-scan planes into subvolumes, which are input to the networks.
The sizes of the subvolumes ($N_A$, $N_B$, and $R$) are {$48\times41\times320$}, {$48\times49\times128$} and {\color{blue}$48\times41\times224$} voxels for the A2A, JHH and {\color{blue}DME} datasets, respectively.
%
For the A2A dataset, we train the networks on three 2080 Ti GPUs with a mini-batch size of 9.
{\color{blue}As to the JHH  and DME datasets, 
we train the networks on one 2080 Ti GPU with the mini-batch sizes of 6 and 4, respectively.
%
%
The networks are trained for 80, 100 and 100 epochs for the A2A, JHH and DME datasets, respectively, with the Adam optimizer \citep{kingma2014adam}.}
The learning rate is set to 0.001 for the A2A dataset;
{\color{blue}for the relatively smaller JHH and DME datasets, the learning rate is initialized to 0.003 and adjusted by a cosine annealing scheduler with the half period and minimum value set to 40 epochs and 3$\times10^{-7}$, respectively.}

For multi-layer segmentation, there are two practical considerations.
First, the set of surface locations $\{\hat{r}_{b,a,l}\}_{l=1}^L$ (where $L$ is the total number of surfaces) predicted for an A-scan are not guaranteed to follow the strict anatomical order.
Therefore, we implement the iterative surface swap trick~\citep{he2019fully}, where the locations of two predicted neighboring surfaces are swapped if they do not obey the correct anatomical order.
Second, {\color{purple}the extents of smoothness of different surfaces vary naturally, and the weight $\lambda$ for the global coherence loss $\mathcal{L}_\mathrm{SmoothS}$ in Eqn. (\ref{eq:L_seg}) should also vary accordingly to accommodate the natural variation.
Empirically, for the $l$\textsuperscript{th} surface,
we compute $\lambda_l=\lambda_b\big/\big({\sum}_{b=1}^{N_B}{\sum}_{a=1}^{N_A}\big\|\triangledown S^g_l(b,a)\big\|\big)$, where $\lambda_b$ is the base weight for a specific dataset and set to 0.1 for the A2A and JHH datasets and 0.03 for the DME dataset, and $S^g_l$ is the ground truth surface.}\footnote{Note that a quick motion correction with NoRMCorre \citep{pnevmatikakis2017normcorre} has to be applied to the A2A dataset beforehand to ensure a valid estimate of $\lambda_l$.}
Intuitively, the smoother the surface naturally is, the more we penalize its global coherence loss.
In practice, we use the arithmetic mean of $\lambda_l$'s of all the training OCT volumes for the $l$\textsuperscript{th} surface.

{\color{purple}For reproducible research, our implementation and trained models are available at: \href{https://github.com/ccarliu/Retinal-OCT-LayerSeg/tree/following-work}{https://github.com/ccarliu/Retinal-OCT-LayerSeg/tree/following-work}.}

\subsection{Motion correction results}

We compare our proposed approach with \citep{montuoro2014motion}, NoRMCorre \citep{pnevmatikakis2017normcorre} and {\color{purple}\citep{fu2016eye}} on both real clinical (the A2A) and synthetic OCT data.
The method proposed by \cite{montuoro2014motion} is based on the hypothesis that a motion-free SD-OCT volume of a healthy person is predominantly locally symmetric along the axial scan direction ($z$ axis), and sequentially corrects the motions in the $z$ and $x$ directions.
%
NoRMCorre is a template matching based algorithm originally proposed for fast and robust motion correction of calcium imaging data, a similar scenario to the B-scan alignment.
Since the B-scans are mostly motion artifact free internally (i.e., no misalignment among the A-scans within a B-scan), we only need to operate NoRMCorre in a rigid fashion.
In addition, given the fast cross-B-scan change in image content due to the low $y$ axial resolution, each B-scan image directly uses its immediate predecessor as the template to match, rather than the median of the buffer as in \citep{pnevmatikakis2017normcorre}.
%
{\color{purple}The method proposed by \citet{fu2016eye} is 
based on the retinal layer saliency map and center bias constraint, to alleviate performance degradation caused by background noise and strong vessels, respectively.}

\begin{figure}[t]
\centering\footnotesize
\includegraphics[width=\columnwidth,trim=0 0 0 0,clip]{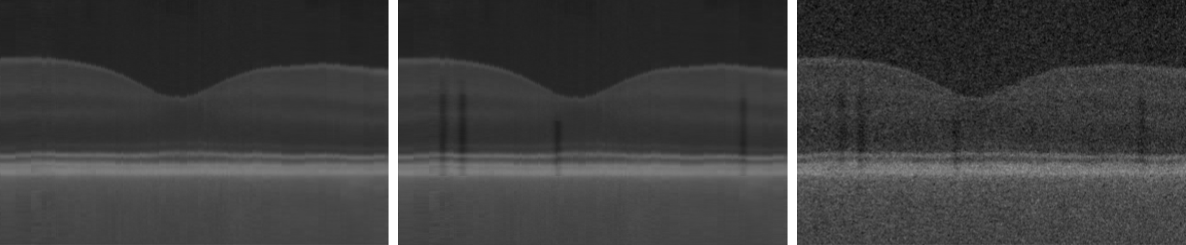}\\
\hspace{.15\columnwidth}(a)\hfill(b)\hfill(c)\hspace*{.15\columnwidth}
\caption{A synthetic B-scan image of an OCT volume created adopting the procedures described in \citep{montuoro2014motion}.
(a) First synthesized without vessel shadow or noise.
(b) Vessel shadows overlaid.
(c) Noise added.
} \label{fig:syn_vis}
\end{figure}

\begin{table}[t]
	\caption{Mean absolute differences (in pixels) between recovered and ground truth motion vectors on the synthetic SD-OCT dataset.
	Results of NoRMCorre \citep{pnevmatikakis2017normcorre}, \citet{montuoro2014motion} and {\color{purple}\citet{fu2016eye}} are presented for comparison.
	Format: mean (std.).}
    \centering
\setlength{\tabcolsep}{1mm}
\begin{adjustbox}{width=\columnwidth}
\begin{tabular}{l|cccc|c}
\hline
Motion     & Montuoro et al. & NoRMCorre   &  {\color{purple}Fu et al.}  & No\_layer            & Ours                 \\ \hline
Axial      & 7.06 (0.25)     & 2.11 (0.53) &   {\color{purple}2.88 (1.32)}   & \textbf{1.76} (0.60) & \textbf{1.76} (0.60) \\
Transverse & 8.92 (6.26)     & 8.64 (5.99) &   {\color{purple}5.91 (2.61)}    & 4.91 (2.26)          & \textbf{4.80} (1.64) \\ \hline
\end{tabular}
\end{adjustbox}
	\label{alignment_syn}
\end{table}

\begin{figure*}[!t]
\footnotesize
\includegraphics[width=\textwidth,trim=0 0 0 0,clip]{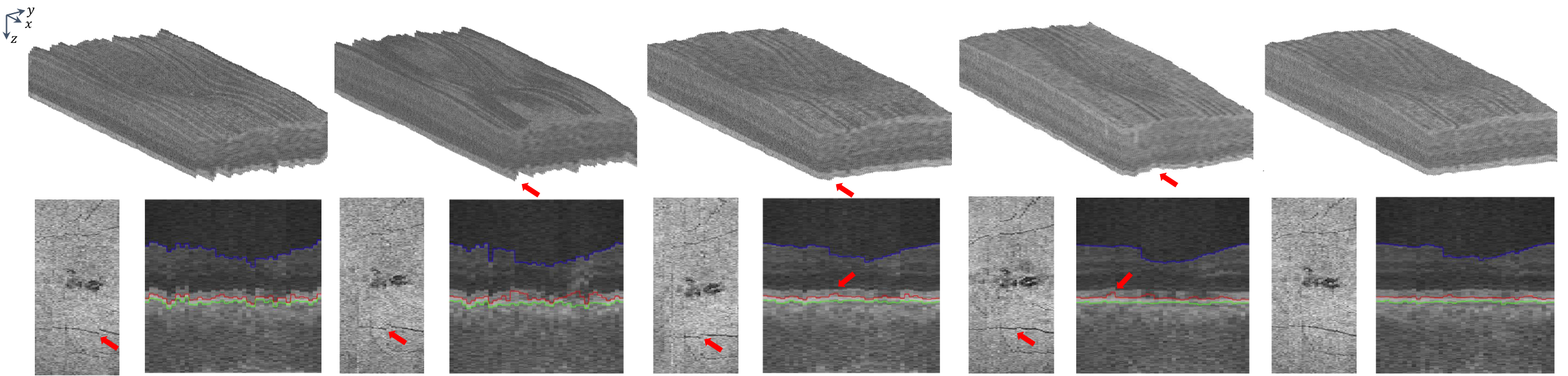}\\
\hspace*{11mm}(a) No correction\hspace{12mm}(b) \cite{montuoro2014motion}\hspace{14mm}(c) NoRMCorre\hspace{18.5mm}{\color{purple}(d) \cite{fu2016eye}}\hspace{16mm}(e) Ours
\caption{Visualization of motion correction results of an OCT volume in the A2A dataset, including: 3D visualization, OCT fundus image obtained by averaging intensity values between the OBM and IRPE surfaces, and $yz$-plane cross-section image (with ground truth layer surfaces overlaid: blue: ILM, red: IRPE, and green: OBM).
NoRMCorre was proposed by \cite{pnevmatikakis2017normcorre}.
Red arrows highlight places where our results are visually better.
} \label{alignment}
\end{figure*}

\subsubsection{Results on synthetic data}

%
For quantitative validation against ground truth motions, we create a synthetic dataset of 20 SD-OCT volumes from the A2A clinical dataset adopting the procedures described in \citep{montuoro2014motion} (Fig. \ref{fig:syn_vis}(a)--(c)).\footnote{\color{yellow}Although the JHH dataset was motion corrected by its provider, we prefer synthetic data here considering potential residual motions of the JHH data.}
Then, we simulate motion artifacts on the synthesized volumes as follows:
(i) for the axial motion in the $z$ axis, each B-scan image is moved by a random displacement uniformly sampled from $[-15,15]$ pixels; and 
(ii) for the transverse motion in the $x$ axis, we group all the B-scans into three to five consecutive groups and apply a random displacement uniformly sampled from $[-15,15]$ pixels to each of the groups as a whole, to simulate the micro-saccades of the eye \citep{fu2016eye}.
%
We repeat the above procedures five times on each of the synthesized volumes, resulting a total of 100 misaligned synthetic volumes.

We then apply various motion correction methods to the purposely misaligned synthetic OCT volumes, to recover the applied motion vectors.
The results are shown in Table \ref{alignment_syn}.
We can see that NoRMCorre substantially outperforms \citep{montuoro2014motion} in axial motion correction by $\sim$70\%, but remains at the same level for transverse motion correction (8.92 versus 8.64 pixels).
{\color{purple}Compared with NoRMCorre, \citep{fu2016eye} substantially improves the transverse motion by $\sim$32\% (5.91 pixels), while slightly increasing the axial residual error.}
In contrast, our method not only further reduces the axial residual errors by apparent advantages (2.11 versus 1.76 pixels), but also substantially reduces the transverse residual errors from 5.91 to 4.80 pixels.
We conjecture this is because 
our method not only relies on the grayscale image information for alignment but additionally fully utilizes the layer segmentation---produced by the coupled segmentation branch.
We implement a variant of our method (No\_layer) where the layer segmentation is not utilized to exclude background for transverse alignment.
As expected, the transverse correction performance drops notably, again emphasizing the value of making use of the segmentation.
{\color{purple}In addition, as we conduct our transverse motion correction as post-processing, it is likely that our method also benefits from its superior performance on axial motion correction for transverse motion correction.}
In conclusion, our proposed method yields the best motion correction performance on the synthetic data.

\begin{table}[t]
\small
\caption{Motion correction results on the A2A clinical SD-OCT dataset, evaluated with the mean absolute distance (MAD; in pixel) of the same surface, and the normalized cross-correlation (NCC) between adjacent B-scans.
 Results of NoRMCorre \citep{pnevmatikakis2017normcorre}, \citet{montuoro2014motion} and {\color{purple}\citet{fu2016eye}} are included for comparison.
	Format: mean (std.).}\label{alignment_a2a}
    \centering
\setlength{\tabcolsep}{.5mm}
\begin{adjustbox}{width=\columnwidth}
\begin{tabular}{rcccc|c}
\hline
\multicolumn{1}{c}{Metric} & No correction        & Montuoro et al.      & NoRMCorre  & {\color{purple}Fu et al.}             & Ours                     \\ \hline
\multicolumn{1}{l}{MAD}    & \multicolumn{1}{l}{} & \multicolumn{1}{l}{}& \multicolumn{1}{l}{} & \multicolumn{1}{l|}{} & \multicolumn{1}{l}{}     \\ \cline{1-1}
ILM                        & 3.92 (1.57)          & 3.42 (1.53)          & 1.74 (0.52)    &   {\color{purple}2.94 (2.15)}    & \textbf{1.58} (0.49)     \\
IRPE                       & 4.17 (1.64)          & 3.68 (1.69)          & 2.19 (0.93)    &   {\color{purple}3.31 (2.22)}    & \textbf{2.12} (0.89)     \\
OBM                        & 3.93 (1.59)          & 3.43 (1.55)          & 1.87 (0.65)    &   {\color{purple}3.06 (2.09)}    & \textbf{1.81} (0.64)     \\
Average                    & 4.00 (1.59)          & 3.51 (1.57)          & 1.93 (0.62)    &   {\color{purple}3.10 (1.58)}    & \textbf{1.83} (0.59)     \\ \hline
\multicolumn{1}{c}{NCC}    & 0.0456 (0.0049)      & 0.0455 (0.0050)      & 0.0470 (0.0067)  &   {\color{purple}0.0483 (0.0068)}   & \textbf{0.0481} (0.0062) \\ \hline
\end{tabular}
\end{adjustbox}
\end{table}

\subsubsection{Results on the A2A dataset}
In addition to the synthetic data, we also evaluate the motion correction performance on the A2A real clinical data.
As the underlying ground truth motions are unknown for the clinical dataset, we instead compute the MAD between the locations of the same surface in adjacent B-scans (note the ground truth surface locations are used here), and the NCC between adjacent B-scans, for quantitative evaluation.
The lower the MAD and the higher the NCC, the better the two B-scans are matched.
The results are shown in Table \ref{alignment_a2a}.
As we can see, all the evaluated motion correction methods improve both metrics upon the baseline before correction.
While the improvements by \citep{montuoro2014motion} are minor, those by NoRMCorre \citep{pnevmatikakis2017normcorre} are more substantial.
{\color{purple}Compared to NoRMCorre, \citep{fu2016eye} is better in NCC but worse in MADs.}
{\color{purple} Our proposed method further improves upon both NoRMCorre and \citep{fu2016eye}, achieving the lowest MADs for all three evaluated surfaces and their average and comparable NCC to \citep{fu2016eye}.}
{\color{purple}Figure \ref{alignment} visualizes motion correction results of an A2A OCT volume by these methods.}
We can observe obvious mis-alignment between the B-scans before correction.
While \citep{montuoro2014motion} hardly realigns the B-scans properly, {\color{purple}NoRMCorre, \citep{fu2016eye} and our method make the B-scans more aligned, and our results are visually better as highlighted by the red arrows}.

\subsection{Layer segmentation results}

\begin{table*}[t]
\caption{Layer segmentation results evaluated by the mean absolute distance ($\upmu$m) between the predicted and ground truth surface locations (std. in parentheses).
Results of ReLayNet \citep{roy2017relaynet}, MGU-Net \citep{li2021multi}, FCBR \citep{he2019fully,he2021structured}, {\color{blue}IPM \citep{xie2022globally}, and DDP \citep{xie2022deep}} are included for comparison. 
The asterisks denote statistically significant differences from our proposed method with the Wilcoxon signed-rank test ($^*$: $p<0.05$; $^{**}$: $p<0.01$; $^{***}$: $p<0.001$).}
\label{tab:MAD}
\centering
\begin{adjustbox}{width=.85\textwidth}
\begin{tabular}{rlllllll}
\hline
\multicolumn{8}{c}{\textit{A2A dataset}}                                                                                                                                                                            \\ \hline
\multicolumn{1}{r|}{Method} & ILM (AMD)            & ILM (Normal)         & IRPE (AMD)           & IRPE (Normal)        & OBM (AMD)            & \multicolumn{1}{c|}{OBM (Normal)}         & Overall              \\ \hline
\multicolumn{1}{r|}{ReLayNet} & 5.23 (10.65)$^{***}$         & 2.44 (4.08)$^{***}$          & 12.45 (24.95)$^{***}$         & 3.80 (3.59)$^{***}$           & 11.55 (19.76)$^{***}$         & \multicolumn{1}{l|}{3.23 (1.83)$^{***}$}           & 7.64 (13.68)$^{***}$          \\
\multicolumn{1}{r|}{MGU-Net}  & 2.48 (4.71)$^{***}$          & 1.49 (0.45)$^{***}$          & 4.51 (7.50)$^{***}$          & 2.39 (1.38)$^{***}$          & 5.22 (4.19)$^{***}$          & \multicolumn{1}{l|}{2.47 (0.34)$^{*}$}          & 3.45 (4.00)$^{***}$          \\
\multicolumn{1}{r|}{FCBR}     & 1.83 (2.74)$^{***}$          & \textbf{1.22} (0.46)$^{***}$ & 3.09 (2.29)          & 2.15 (1.38)  & 4.51 (3.28)           & \multicolumn{1}{l|}{\textbf{2.28} (0.34)$^{***}$} & 2.74 (1.87)$^{*}$          \\ 
\multicolumn{1}{r|}{{\color{blue}IPM}}     & \color{blue}\textbf{1.80} (1.83)          & \color{blue}1.28 (0.41)$^{*}$ & \color{blue}3.15 (1.87)$^{***}$          & \color{blue}2.18 (1.26)$^{*}$  &\color{blue} 4.46 (2.46)           & \multicolumn{1}{l|}{\color{blue}2.31 (0.34)$^{***}$} &\color{blue} 2.75 (1.66)$^{*}$          \\ 
\multicolumn{1}{r|}{{\color{blue}DDP}}     & \color{blue}1.81 (2.38)         & \color{blue}1.23 (0.46)$^{*}$ & \color{blue}3.11 (2.09)$^{*}$           & \color{blue}2.12 (1.25)  & \color{blue}4.46 (3.63)           & \multicolumn{1}{l|}{\color{blue}2.31 (0.38)$^{***}$} & \color{blue}2.73 (1.48)$^{*}$          \\ \hline
\multicolumn{1}{r|}{Proposed} & \textbf{1.80} (1.97) & 1.30 (0.52)          & \textbf{2.91} (1.61) & \textbf{2.10} (1.35) & \textbf{4.34} (2.55) & \multicolumn{1}{l|}{2.40 (0.38)}          & \textbf{2.68} (1.39) \\ \hline
\end{tabular}
\end{adjustbox}
\setlength{\tabcolsep}{1mm}
\begin{adjustbox}{width=\textwidth}
\begin{tabular}{r|lllllllll|l}
\multicolumn{11}{c}{\textit{JHH dataset}} \\
\hline
Method & ILM                  & RNFL-GCL             & IPL-INL              & INL-OPL             & OPL-ONL              & ELM                 & IS-OS                & OS-RPE               & BM                   & Overall              \\ \hline
ReLayNet & 2.67 (0.46)$^{***}$           & 3.58 (1.11)$^{***}$           & 3.41 (0.81)$^{***}$           & 3.42 (0.67)$^{***}$          & 3.11 (0.82)$^{*}$           & 2.96 (0.74)$^{**}$          & 2.43 (0.95)$^{**}$           & 4.02 (1.15)$^{**}$          & 3.45 (2.05)$^{**}$           & 3.23 (0.71)$^{***}$           \\
MGU-Net  & 2.58 (0.26)$^{***}$           & 3.16 (0.64)$^{***}$           & 3.02 (0.38)$^{***}$        & 3.29 (0.53)          & 2.90 (0.58)$^{***}$           & 2.82 (0.69)$^{*}$          & 2.15 (0.59)           & 3.60 (0.62)          & 3.59 (2.34)$^{**}$           & 3.01 (0.42)$^{***}$           \\
FCBR     & 2.39 (0.30)$^{*}$           & 3.01 (0.70)$^{***}$           & 2.96 (0.38)$^{**}$           & 3.24 (0.51) & 2.86 (0.55)$^{***}$           & 2.71 (0.86) & 1.98 (0.75) & 3.52 (0.94) & \textbf{2.74} (1.67)$^{*}$ & 2.82 (0.40)           \\
{\color{blue}IPM}     & \color{blue}2.32 (0.48)$^{*}$           & \color{blue}2.94 (0.68)$^{***}$           & \color{blue}2.92 (0.38)$^{**}$           &\color{blue} \textbf{3.15} (0.36) & \color{blue}2.77 (0.57)$^{**}$           & \color{blue}2.73 (0.87) & \color{blue}2.01 (0.76) & \color{blue}3.46 (0.92) & \color{blue}3.04 (2.13) & \color{blue}2.81 (0.48)           \\
{\color{blue}DDP}    & \color{blue}2.32 (0.26)$^{*}$           & \color{blue}3.10 (0.65)$^{***}$           & \color{blue}2.94 (0.36)$^{**}$           & \color{blue}3.17 (0.49) & \color{blue}2.74 (0.55)$^{**}$           & \color{blue}\textbf{2.61} (0.63) & \color{blue}\textbf{1.94} (0.66) & \color{blue}\textbf{3.31} (0.80)$^{*}$ & \color{blue}2.95 (1.97) & \color{blue}2.79 (0.41)           \\
\hline
Proposed & \textbf{2.21} (0.35) & \textbf{2.73} (0.61) & \textbf{2.79} (0.42) & 3.18 (0.33)          & \textbf{2.62}  (0.58) & 2.65 (0.52)          & 2.04 (0.73)           & 3.56 (1.04)           & 3.19 (2.02)           & \textbf{2.77}  (0.51) \\ \hline
\end{tabular}
\end{adjustbox}
\end{table*}

\begin{table*}[t]
\color{blue}
\caption{\color{blue} Layer segmentation results evaluated by the mean 95\textsuperscript{th} percentile of the Hausdorff distance ($\upmu$m) between the predicted and ground truth surface locations (std. in parentheses).
Results of ReLayNet \citep{roy2017relaynet}, MGU-Net \citep{li2021multi}, FCBR \citep{he2019fully}, IPM \citep{xie2022globally}, and DDP \citep{xie2022deep} are included for comparison.
The asterisks denote statistically significant differences from our proposed method with the Wilcoxon signed-rank test ($^*$: $p<0.05$; $^{**}$: $p<0.01$; $^{***}$: $p<0.001$).
}\label{tab:hd95result}
\centering
\begin{adjustbox}{width=.85\textwidth}
\begin{tabular}{rlllllll}
\hline
\multicolumn{8}{c}{\textit{A2A dataset}}                                                                                                                                                                            \\ \hline
\multicolumn{1}{r|}{Method} & ILM (AMD)            & ILM (Normal)         & IRPE (AMD)           & IRPE (Normal)        & OBM (AMD)            & \multicolumn{1}{c|}{OBM (Normal)}         & Overall              \\ \hline
\multicolumn{1}{r|}{ReLayNet} & 19.31 (39.37)$^{***}$         & 8.29 (19.52)$^{***}$          & 42.90 (61.92)$^{***}$         & 13.97 (18.57)$^{***}$           & 34.98 (45.63)$^{***}$         & \multicolumn{1}{l|}{10.07 (10.37)$^{***}$}           & 25.49 (37.79)$^{***}$          \\
\multicolumn{1}{r|}{MGU-Net}  & 6.96 (12.59)$^{***}$ & 3.98(2.64)$^{***}$          & 15.97 (26.35)$^{*}$          & 5.86 (2.81)$^{**}$          & 15.32 (14.30)$^{*}$          & \multicolumn{1}{l|}{5.95 (0.89)$^{*}$}            & 10.36 (13.22)$^{**}$          \\
\multicolumn{1}{r|}{FCBR}     & 4.32 (7.71)$^{***}$          & \textbf{2.68} (1.34)$^{***}$ & 8.55 (8.29)          & 4.75 (2.91)  & 11.43 (9.56)           & \multicolumn{1}{l|}{\textbf{5.00} (0.93)$^{***}$} & 6.82 (5.62)$^{**}$          \\ 
\multicolumn{1}{r|}{IPM}     &  \textbf{4.12} (4.90)$^{**}$          & 3.02 (1.43)$^{***}$ & 9.05 (7.51)$^{***}$          & 4.87 (2.54)$^{*}$  & 11.20 (7.93)           & \multicolumn{1}{l|}{5.04 (0.87)$^{***}$} & 6.91 (4.82)$^{***}$          \\ 
\multicolumn{1}{r|}{DDP}     & 4.45 (6.22)$^{***}$         & 2.95 (1.36)$^{***}$ & 8.92 (7.54)$^{***}$           & 4.97 (2.64)$^{***}$  & 12.83 (15.87)           & \multicolumn{1}{l|}{5.09 (0.89)$^{**}$} & 7.33 (7.02)$^{*}$          \\ \hline
\multicolumn{1}{r|}{Proposed} & 4.37 (5.88) & 2.89 (1.54)          & \textbf{7.92} (5.80) & \textbf{4.68} (2.86) & \textbf{10.72} (7.56) & \multicolumn{1}{l|}{5.28 (1.00)}          & \textbf{6.58} (4.21) \\ \hline
\end{tabular}
\end{adjustbox}

\begin{adjustbox}{width=\textwidth}
\begin{tabular}{r|ccccccccc c}
\multicolumn{10}{c}{\textit{JHH dataset}} \\
\hline
Method & ILM                  & RNFL-GCL             & IPL-INL              & INL-OPL             & OPL-ONL              & ELM                 & IS-OS                & OS-RPE               &\multicolumn{1}{l|}{ BM}               & Overall                  \\ \hline
ReLayNet & 6.24 (0.95)$^{***}$          & 11.02 (4.79)$^{***}$          & 9.34 (3.14)$^{***}$        & 9.15 (3.05)$^{*}$         & 9.18 (3.62)$^{***}$          & 7.54 (2.92)$^{**}$         & 5.46 (1.53)$^{***}$          & 9.65(3.83)$^{**}$         &\multicolumn{1}{l|}{ 8.08 (3.14)$^{***}$ }     &  8.41 (2.44)$^{***}$   \\
MGU-Net  & 6.03 (0.64)$^{***}$         & 8.96 (2.44)$^{***}$           & 7.86 (1.13)$^{**}$        & 7.98 (1.22)        & 7.96 (1.94)$^{***}$         & 6.45 (1.42)$^{**}$          &4.90 (1.02)$^{*}$       & 7.98 (1.18)      & \multicolumn{1}{l|}{6.90 (3.16)$^{**}$}        &  7.23 (0.98)$^{***}$    \\
FCBR     & 5.52 (0.66)        & 8.38 (2.26)$^{***}$         & 7.68 (1.06)$^{**}$       & 7.90 (1.18) & 7.70 (1.81)         & 6.09 (1.59)$^{**}$ & 4.50 (1.48) & 7.61 (1.64) & \multicolumn{1}{l|}{\textbf{5.61} (2.55)$^{**}$} & 6.78 (0.97) \\
IPM     & 5.49 (0.67)         & 8.34 (2.01)$^{***}$      & 7.66 (0.80)$^{**}$           & 7.95 (0.98) & 7.71 (1.70)$^{*}$      & 6.15 (1.19) & 4.59 (1.50) & 7.69 (1.71) & \multicolumn{1}{l|}{6.07 (3.00) } & 6.84 (0.95)  \\
DDP    & 5.43 (0.74)      & 8.66 (2.17)$^{***}$         & 7.47 (0.90)  & \textbf{7.80} (1.08)         & 7.63 (1.84)$^*$ & \textbf{5.94} (1.31) &\textbf{4.35} (1.12) & \textbf{7.37} (1.54)$^*$ & \multicolumn{1}{l|}{5.96 (2.87)}   & \textbf{6.73} (0.92)      \\ \hline
Proposed & \textbf{5.32} (0.74) & \textbf{7.64} (1.94) & \textbf{7.30} (0.93) & 7.86 (0.84)      & \textbf{7.39} (1.78) & 6.00 (1.17)         & 4.61 (1.25)          & 8.18 (2.23)        & \multicolumn{1}{l|}{6.45 (2.98)}    & 6.75 (1.11)   \\ \hline
\end{tabular}
\end{adjustbox}
\end{table*}

\subsubsection{Comparison with state-of-the-art (SOTA) methods}

We compare our proposed method with several up-to-date baselines: ReLayNet, MGU-Net, FCBR, {\color{blue}IPM, and DDP}. 
ReLayNet \citep{roy2017relaynet} is a U-Net based method which outputs the layer maps.
MGU-Net \citep{li2021multi} employs graph convolutional networks  to simultaneously label the retinal layers and optic disc.
FCBR \citep{he2019fully,he2021structured} is a SOTA method implementing 2D surface regression, thus can directly output surface locations like ours.
{\color{blue}IPM and DDP further employ the primal-dual interior-point method (IPM) and differentiable dynamic programming (DDP) to explicitly enforce surface interaction and smoothness constraints, respectively.}
For the methods that only output layer maps (ReLayNet and MGU-Net), we obtain the surface locations by summing up the output layer maps in each A-scan as done by \cite{he2021structured}. 
We use the official implementation of MGU-Net and RelayNet, and implement and empirically optimize FCBR, {\color{blue}IPM, and DDP}, to get their results.

\begin{figure*}[t]
\centering
\includegraphics[width=0.9\textwidth,trim=0 0 0 0,clip]{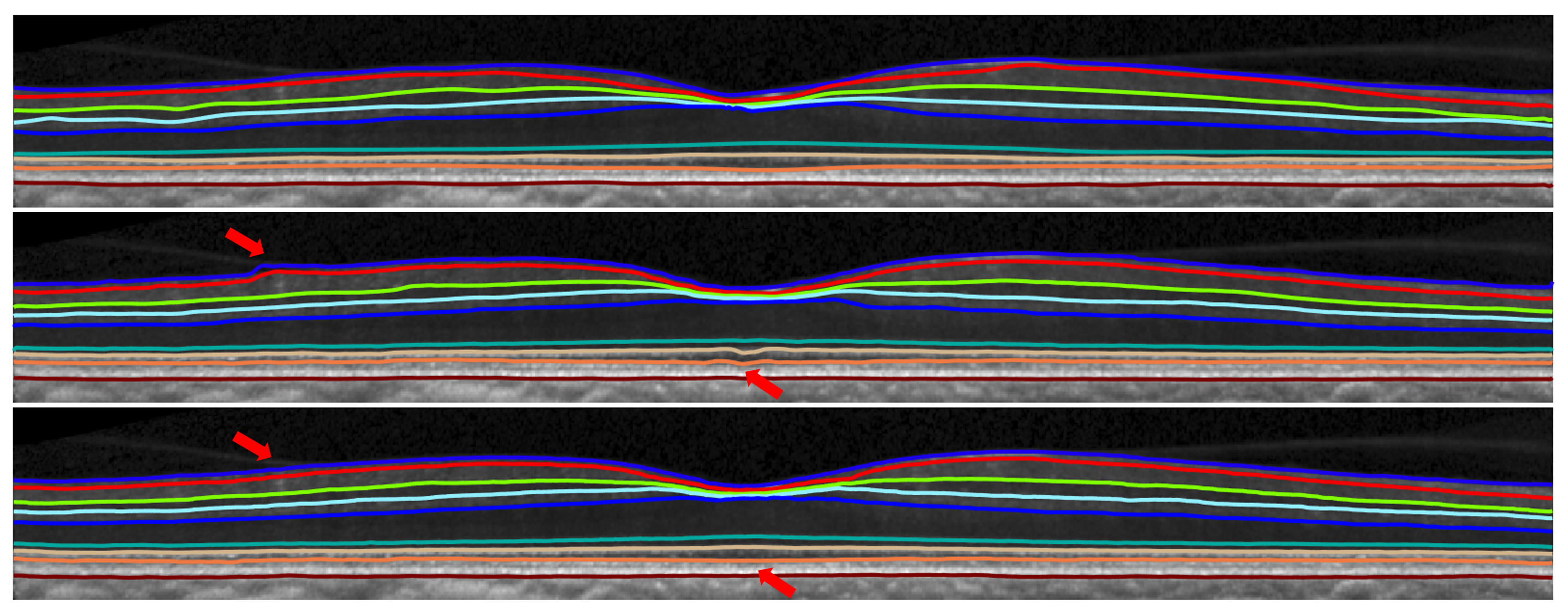}
\caption{Visualization of the manual segmentation (top), and segmentations by FCBR \citep{he2019fully} (middle) and our framework (bottom) of a B-scan (an MS case) in the JHH dataset.
Layer boundaries from top to bottom: ILM, RNFL-GCL, IPL-INL, INL-OPL, OPL-ONL, ELM, IS-OS, OS-RPE, and BM surfaces.
The red arrows indicate where our segmentation is better.
{\color{purple}Note that our framework correctly segments the foveal pit region.}
} \label{fig:seg_JHH}
\end{figure*}

\begin{figure*}[t]
    \centering
    \includegraphics[width=\textwidth]{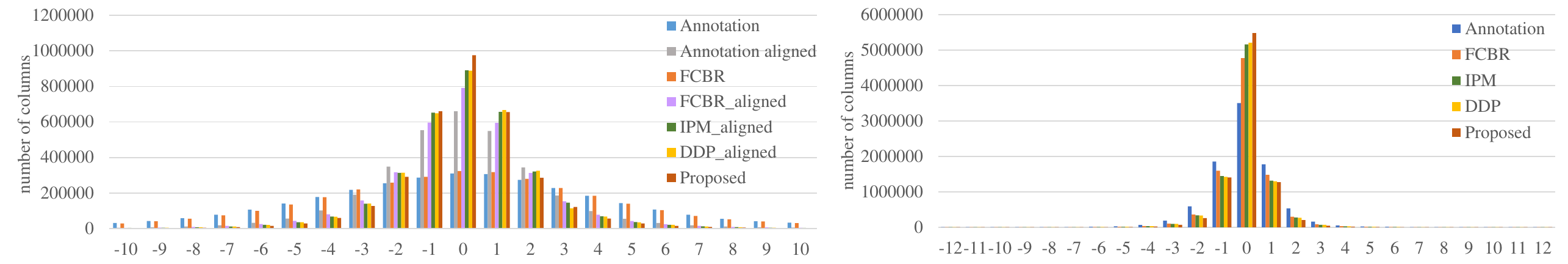}
    \caption{\color{purple}Histograms of the surface distances ($x$ axis; in pixels) between adjacent B-Scans.
    Left: A2A dataset, and right: JHH dataset.}
    \label{fig:histogram}
\end{figure*}

\begin{figure}[t]
\centering
\includegraphics[width=\columnwidth,trim=0 0 0 0,clip]{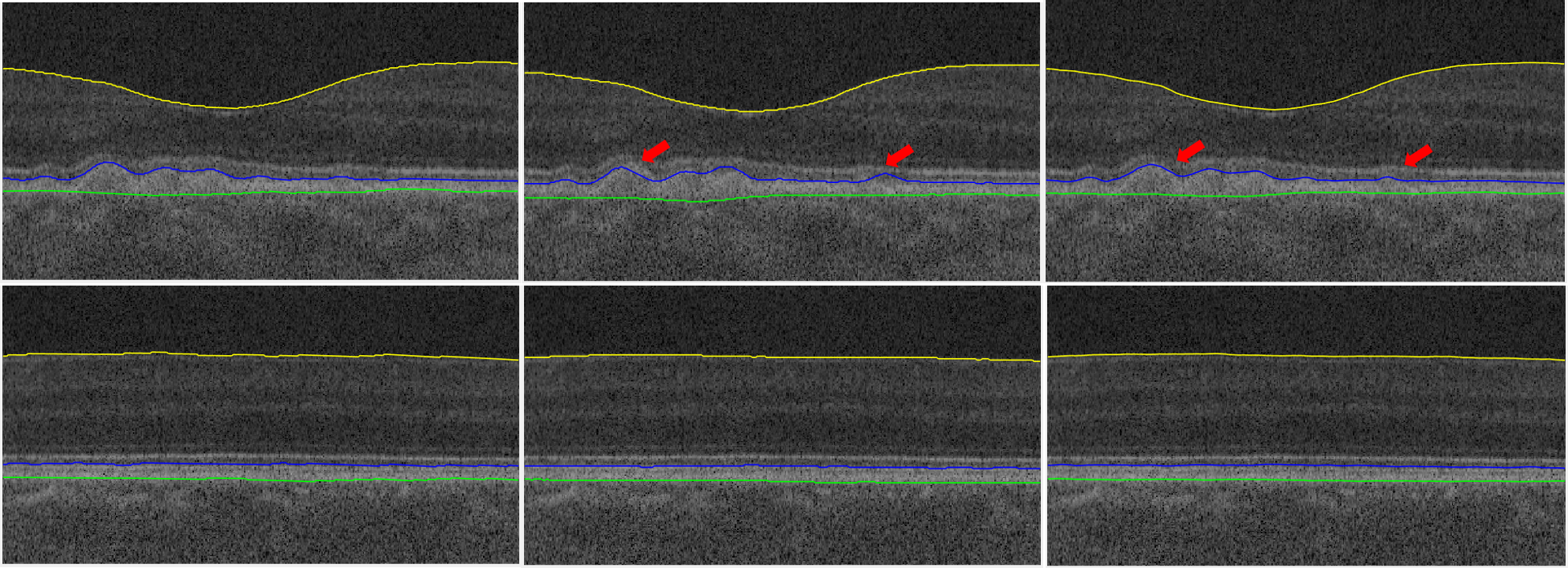}
\caption{Example manual segmentation (left), and segmentation by FCBR \citep{he2019fully} (middle) and our framework (right) on the A2A dataset.
Top: an AMD case; bottom: a normal control.
The yellow, blue, and green curves indicate the ILM, IRPE, and OBM boundaries, respectively. 
The red arrows indicate where our segmentation is better.
{\color{purple}Note that our framework correctly segments the AMD case in the presence of drusen.}} \label{fig:seg_A2A}
\end{figure}

\begin{figure}[t]
    \centering
    \includegraphics[width=.45\textwidth]{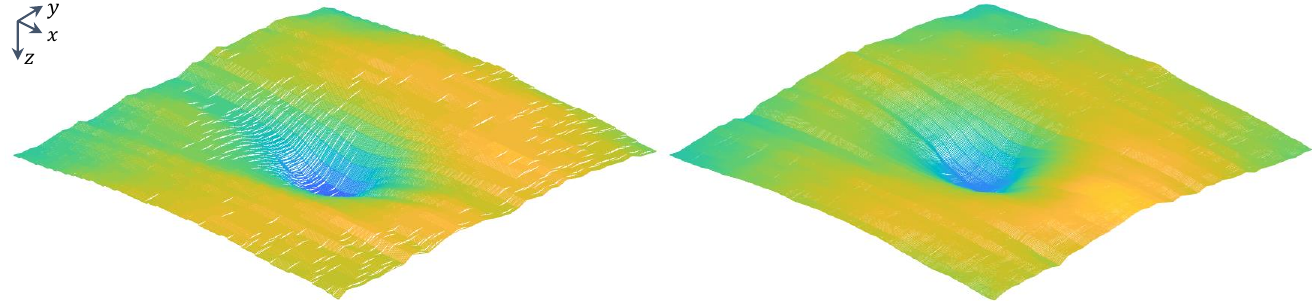}
    \caption{\color{purple}3D surface visualization of the segmented ILM layer of an OCT volume in the A2A dataset.
    Left: FCBR with B-scan pre-alignment by NoRMCorre, and right: our method.}
    \label{fig:surf}
\end{figure}

The MADs between the predicted surface locations and manual delineation on the A2A and JHH SD-OCT datasets are charted in Table \ref{tab:MAD}.
{\color{blue}On the A2A dataset, our method achieves a significantly lower overall MAD with a smaller standard deviation (2.68$\pm$1.39 $\upmu$m) than the previous best performing methods: FCBR (2.74$\pm$1.87 $\upmu$m), IPM (2.75$\pm$1.66 $\upmu$m), and DDP (2.73$\pm$1.48 $\upmu$m), all with $p<0.05$.}
Meanwhile, 
our method, FCBR, IPM, and DDP are substantially better than the other two compared methods ReLayNet (7.64$\pm$13.68 $\upmu$m) and MGU-Net (3.45$\pm$4.00 $\upmu$m) by large margins.
On the JHH dataset, we note that the performance variations between the methods largely decrease,
probably because this dataset does not present severe pathologies and is better in image quality, thus is easier to segment than the A2A dataset.
{\color{blue}Our method achieves the lowest overall MAD of 2.77 $\upmu$m, marginally lower than that of DDP (2.79 $\upmu$m), IPM (2.81 $\upmu$m) and FCBR (2.82 $\upmu$m) with no statistical significance.}
Notwithstanding, our overall MAD is still significantly better than that of ReLayNet (3.23 $\upmu$m) and MGU-Net (3.01 $\upmu$m), both with $p<0.001$.

{\color{blue}The mean HD95 values between the predicted surfaces and manual delineation on the A2A and JHH datasets are presented in Table \ref{tab:hd95result}.
The general trends are the same as the MADs in Table \ref{tab:MAD}.
On the A2A dataset, our method is significantly better than all others in term of the overall mean HD95, although the performance gaps between the methods become more obvious using HD95 as the evaluation metric.
On the JHH dataset, the performance of our method (overall mean HD95: 6.75 $\upmu$m) is comparable to that of FCBR and DDP (6.78 and 6.73 $\upmu$m, respectively, no statistical significance), slightly better than that of IPM (6.84 $\upmu$m, no statistical significance), while significantly better than that of ReLayNet and MGU-Net (8.41 and 7.23 $\upmu$m, respectively, $p<0.001$).}

Figs. \ref{fig:seg_JHH} and \ref{fig:seg_A2A} show example segmentation by our framework and FCBR on the JHH and A2A datasets, respectively.
In most cases the segmentation by both methods looks comparable, yet in difficult situations (pointed by red arrows) our framework produces layer boundaries closer to the manual segmentation.

\begin{figure}[t]
\centering
\includegraphics[width=\columnwidth,trim=0 0 0 0,clip]{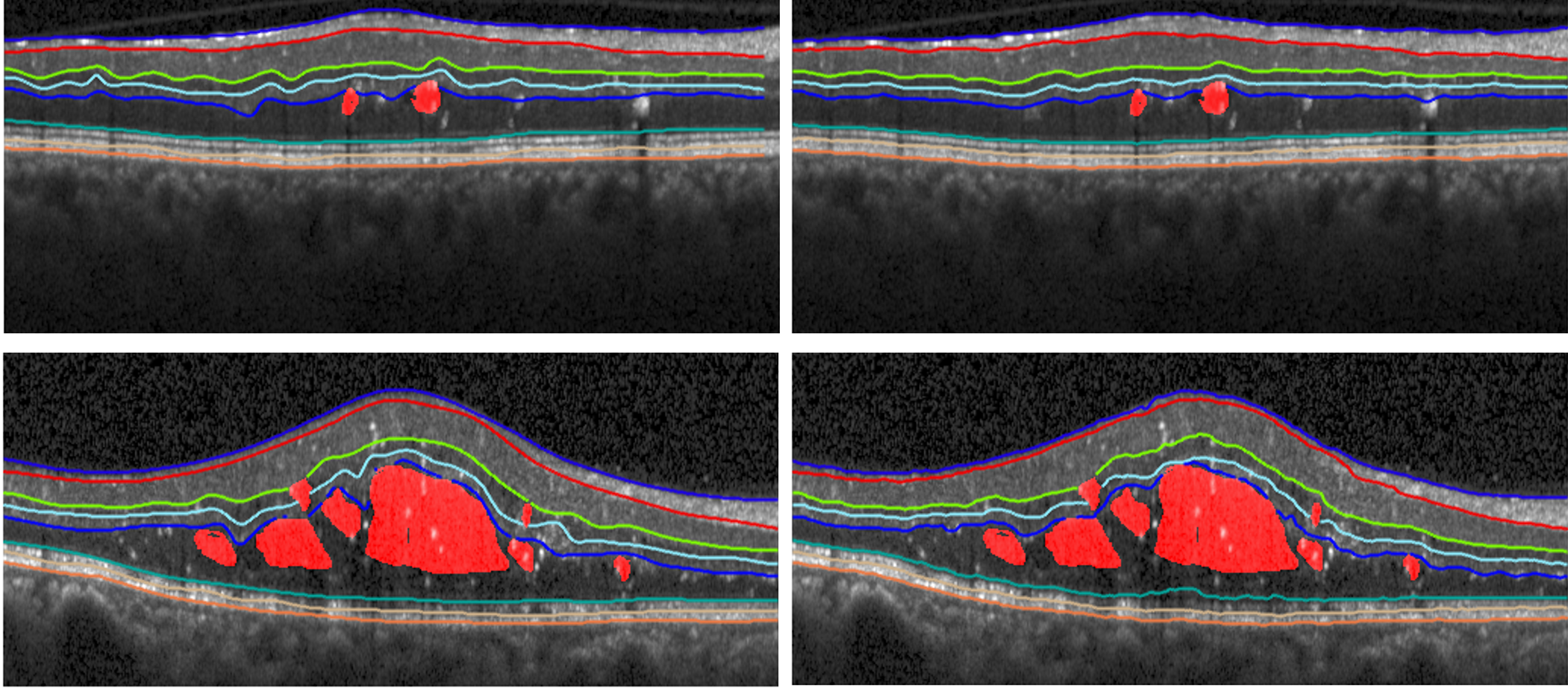}
\caption{\color{purple}Example manual segmentation (left) and segmentation by our framework (right) of two B-scans in the Duke DME dataset.
Layer boundaries from top to bottom: ILM, RNFL-GCL, IPL-INL, INL-OPL, OPL-ONL, IS-OS, OS-RPE, and BM surfaces.
The red area indicates edema.
Note that our framework correctly segments the layers in the presence of modest and severe pathologies.} \label{fig:seg_DME}
\end{figure}

\subsubsection{B-Scan connectivity analysis}
A hypothesized advantage of our proposed 3D OCT layer segmentation over 2D counter-methods such as FCBR \citep{he2019fully,he2021structured} is the cross-B-scan surface smoothness, i.e., 3D continuity beyond the 2D B-scan image planes.
To test the hypothesis, we compute the surface distance between adjacent B-scans by $\big|r_{b+1,a}-r_{b,a}\big|$ to quantify the cross-B-scan (dis)continuity.
The surface distance histograms for the A2A and JHH datasets are shown in Fig. \ref{fig:histogram}.
{\color{purple}On the A2A dataset, the surfaces segmented by our framework have better cross-B-scan connectivity than those by FCBR, as indicated by the more conspicuous spikes clustered around 0 of our framework.
After B-scan pre-alignment by NoRMCorre \citep{pnevmatikakis2017normcorre}, the connectivity of FCBR improves, yet is still inferior to that of our framework.
For intuitive perception, we visualize the ILM layer segmented by FCBR (with NoRMCorre pre-alignment) and our framework in Fig. \ref{fig:surf}.
It can be observed that our segmentation is visually smoother.
Similarly, IPM \citep{xie2022globally} and DDP \citep{xie2022deep} with pre-alignment still lag behind our framework on connectivity.
This suggests that merely conducting 3D alignment does not guarantee 3D continuity of the segmentation results, as long as the B-scans are handled separately.}
It is worth noting that our method also achieves better cross-B-scan connectivity than the ground truth after alignment, likely due to the same reason (i.e., human annotators work with one B-scan at a time).
{\color{purple}On the JHH dataset, the surfaces segmented by our framework again show the best cross-B-scan connectivity, as expected} (note that the JHH data are already motion-corrected by the provider, thus we do not preprocess them with motion correction for compared methods or manual segmentation).

\begin{table*}[t]
\caption{Ablation study results evaluated by the mean absolute distance ($\upmu$m) between the predicted and ground truth surface locations (std. in parentheses).
{\color{blue}The asterisks denote statistically significant differences from our proposed full model with the Wilcoxon signed-rank test ($^*$: $p<0.05$; $^{**}$: $p<0.01$; $^{***}$: $p<0.001$).}
}
\label{tab:abalate}
\centering

\begin{adjustbox}{width=.85\textwidth}
\begin{tabular}{c|cccccc|c}
\hline
\multicolumn{8}{c}{\textit{A2A dataset}}\\
\hline
Ablation   & ILM (AMD)            & ILM (Normal)         & IRPE (AMD)           & IRPE (Normal)        & OBM (AMD)            & OBM (Normal)         & Overall              \\ \hline
no\_align  & 2.25 (3.77)$^{***}$           & 1.40 (0.42)           & 3.14 (1.72)$^{***}$           & 2.18 (1.37)$^{*}$           & 4.96 (3.26)$^{***}$           & 2.49 (0.40)$^{**}$           & 3.00 (1.78)$^{***}$           \\
pre\_align & 1.80 (2.36)           & 1.30 (0.49)           & 3.09 (1.79)$^{***}$           & \textbf{2.05} (1.40) & 4.75 (3.61)$^{*}$           & \textbf{2.34} (0.37)$^{*}$ & 2.77 (1.79)$^{*}$           \\
\color{blue}cascade  & \color{blue}1.73 (2.13)$^{**}$ & \color{blue}\textbf{1.26} (0.45)$^{**}$ & \color{blue}3.08 (1.98)$^{*}$ & \color{blue}2.16 (1.35)$^{**}$ & \color{blue}4.53 (3.02) & \color{blue}2.45 (0.40)$^{**}$ & \color{blue}2.74 (1.56)$^{*}$ \\
\color{blue}no\_smooth &\color{blue} \textbf{1.68} (1.84) &\color{blue} 1.27 (0.47)           &\color{blue} 3.10 (1.97)$^{***}$           &\color{blue} 2.13 (1.45)$^{*}$           &\color{blue} 4.84 (3.43)$^{***}$           &\color{blue} 2.45 (0.41)$^{*}$           & \color{blue}2.81 (1.53)           \\
3D-3D      & 1.87 (2.19)$^{*}$           & 1.31 (0.46)$^{*}$           & 3.12 (1.74)$^{*}$           & 2.13 (1.45)$^{***}$           & 4.78 (2.99)$^{***}$           & 2.43 (0.40)$^{*}$           & 2.85 (1.82)$^{**}$           \\ \hline
Proposed & 1.80 (1.97) &  1.30 (0.52)          & \textbf{2.91} (1.61) & 2.10 (1.35) & \textbf{4.34} (2.55) & \multicolumn{1}{c|}{2.40 (0.38)}          & \textbf{2.68} (1.39) \\  \hline
\end{tabular}
\end{adjustbox}
\begin{adjustbox}{width=\textwidth}
\begin{tabular}{c|ccccccccc|c}
\multicolumn{11}{c}{\textit{JHH dataset}} \\
\hline
Ablation   & ILM                  & RNFL-GCL             & IPL-INL              & INL-OPL    & OPL-ONL              & ELM        & IS-OS                & OS-RPE               & BM                   & Overall              \\ \hline
no\_align  & 2.29 (0.47)           & 2.75 (0.64)           & 2.92 (0.37)           & 3.20 (0.45) & 2.84 (0.68)           & \textbf{2.65} (0.76) & \textbf{1.94} (1.02) & \textbf{3.39} (0.82) & 3.09 (2.07)           & 2.79 (0.57)           \\
no\_smooth & 2.31 (0.36)$^{*}$           & 2.85 (0.70)$^{*}$           & 2.91 (0.56)           & 3.26 (0.39)$^{**}$ & 2.70 (0.59)$^{**}$           & 2.73 (0.61) & 1.99 (0.55)           & 3.54 (1.07)           & \textbf{2.90} (2.02) & 2.80 (0.49)           \\
3D-3D      & 2.34 (0.35)$^{*}$           & 2.88 (0.73)$^{**}$           & 3.03 (0.56)$^{**}$           & 3.50 (0.52)$^{**}$ & 2.80 (0.63)$^{**}$           & \textbf{2.65} (0.58) & 1.99 (0.95)           & 3.59 (0.90)           & 3.11 (1.58)           & 2.88 (0.51)$^{*}$           \\ \hline
Proposed   & \textbf{2.21} (0.35) & \textbf{2.73} (0.61) & \textbf{2.79} (0.42) & \textbf{3.18} (0.33)          & \textbf{2.62}  (0.58) & \textbf{2.65} (0.52)          & 2.04 (0.73)           & 3.56 (1.04)           & 3.19 (2.02)           & \textbf{2.77}  (0.51) \\ \hline
\end{tabular}
\end{adjustbox}

\end{table*}

\begin{figure*}[t]
\centering
\includegraphics[width=.45\textwidth]{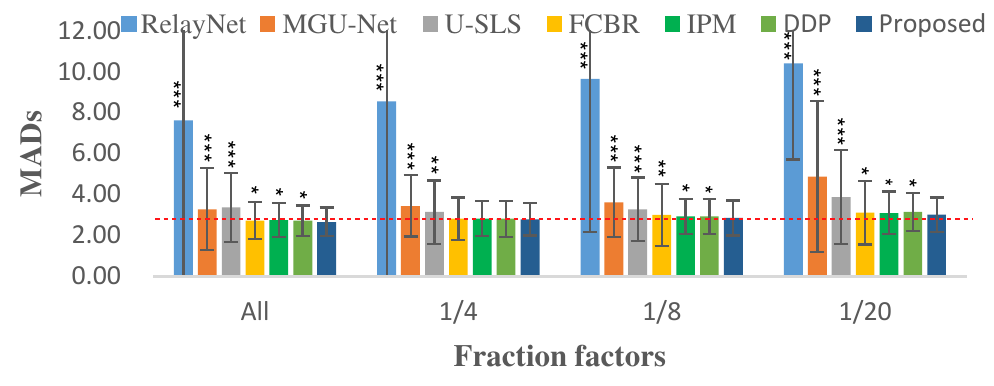}\phantom{W}\includegraphics[width=.52\textwidth]{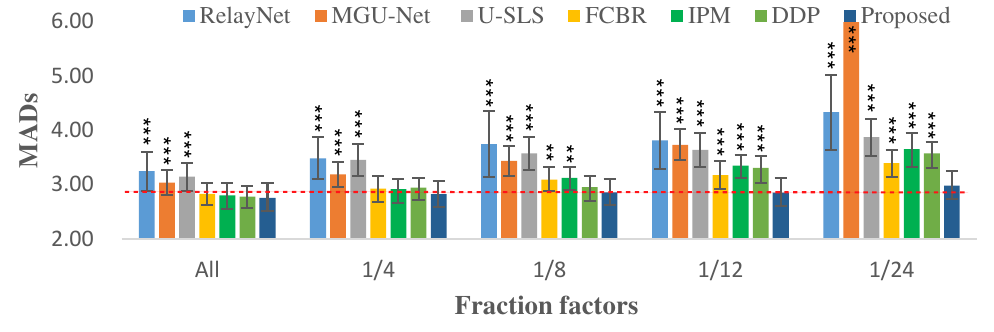}
\caption{\color{purple}Semi-supervised layer segmentation results in overall MAD ($\upmu$m) on the A2A (left) and JHH (right) datasets with different fraction factors (standard deviation overlaid).
The dashed horizontal lines indicate the performance of our proposed method with the 1/8 fraction factor.
The asterisks denote statistically significant differences from our proposed method with the Wilcoxon signed-rank test ($^*$: $p<0.05$; $^{**}$: $p<0.01$; $^{***}$: $p<0.001$).
} \label{semi_result_a2a}
\end{figure*}

\begin{table*}[t]
    
    \caption{\color{blue}Layer segmentation results evaluated by the mean absolute distance ($\upmu$m) between the predicted and ground truth surface locations on the Duke DME dataset {\color{purple}(std. in parentheses)}.
    Results of FCBR \citep{he2019fully,he2021structured} and three SOTA graph-based methods, i.e., \cite{Chiu:15}, \cite{karri2016learning}, and \cite{rathke2017locally}, are included for comparison.
    {\color{purple}\textit{Note:} results of comparing methods are directly cited from \cite{he2019fully,he2021structured} based on the same data split and evaluation protocol, with no standard deviation reported though.}
    }\label{tab:dmeresult}
    \centering\color{blue}
    \begin{adjustbox}{width=.9\textwidth}
    \begin{tabular}{c|cccccccc|c}
\hline
Method                                                           & ILM                         & RNFL-GCL                             & IPL-INL                              & INL-OPL                              & OPL-ONL                              & IS-OS                                & OS-RPE                               & BM                                   & Overall                              \\ \hline
\cite{Chiu:15}                                                   & 6.59                        & 8.38                                 & 9.04                                 & 11.02                                & 11.01                                & 4.84                                 & 5.74                                 & 5.91                                 & 7.82                                 \\
\cite{karri2016learning}                                         & \textbf{4.47}               & 11.77                                & 11.12                                & 17.54                                & 16.74                                & 4.99                                 & 5.35                                 & 4.30                                 & 9.54                                 \\
\cite{rathke2017locally}                                         & 4.66                        & 6.78                                 & 8.87                                 & 11.02                                & 13.60                                & 4.61                                 & 8.06                                 & 5.11                                 & 7.71                                 \\
FCBR \citep{he2019fully,he2021structured} & 4.51 & \textbf{6.71} & 8.29          & 10.71         & 9.88          & \textbf{4.41} & 4.52 & 4.61 & 6.70          \\ \hline
{\color{purple} Ours}                                      & {\color{purple} 4.62 (0.63)} & {\color{purple} 7.51 (2.13)}          & {\color{purple} \textbf{7.52 (1.73)}} & {\color{purple} \textbf{8.49 (2.09)}} & {\color{purple} \textbf{9.40 (2.49)}} & {\color{purple} 4.75 (1.16)}          & {\color{purple} \textbf{4.50 (0.50)}}          & {\color{purple} \textbf{4.16 (0.50)}}          & {\color{purple} \textbf{6.37 (1.01)}} \\ \hline
\end{tabular}
    \end{adjustbox}
    
\end{table*}

\subsubsection{Ablation study}
Next, we conduct ablation experiments to verify the effectiveness of the design and each module of the proposed framework.
Specifically, we evaluate several variants of our model: no\_align (without the alignment branch or pre-alignment), {\color{purple}pre\_align (without the alignment branch but pre-aligned by NoRMCorre \citep{pnevmatikakis2017normcorre})}, {\color{blue}cascade (cascading an alignment network and a pure 3D encoder-decoder segmentation network; 
or in other words, breaking our model into two cascading, exclusive networks for B-scan alignment and 3D segmentation, respectively)}, no\_smooth (without the global coherence loss $\mathcal{L}_\mathrm{SmoothS}$), and 3D-3D (replacing the encoder $G_f$ with 3D CNNs).
The results on the A2A dataset are presented in Table \ref{tab:abalate} top, from which several conclusions can be drawn.
First, the variant without any alignment (no\_align) yields the worst results, suggesting that the mismatch between B-scans does have a negative impact on 3D analysis of OCT data such as the 3D surface segmentation.
Second, our full model integrating the alignment branch improves over both pre\_align {\color{blue}and cascade}. 
We speculate this is because the alignment branch can produce better alignment results than pre\_align, and more importantly, {\color{blue}it produces a slightly different alignment each time, serving as a kind of data and feature augmentation for enhanced diversity for the segmentation decoder $G_s$.}
Third, removing $\mathcal{L}_\mathrm{SmoothS}$ (no\_smooth) apparently decreases the performance, demonstrating its effectiveness in exploiting the anatomical prior of smoothness.
Lastly, our hybrid 2D-3D framework outperforms its counterpart 3D-3D network, indicating that the 2D CNNs can better deal with the mismatched B-scans prior to proper realignment.

The ablation results on the JHH dataset are shown in Table \ref{tab:abalate} bottom.
Since this dataset was acquired on a scanner with built-in motion correction,
we do not evaluate the pre\_align {\color{blue}or cascading} variant on it.
As can be seen, 
our full model still slightly outperforms the no\_align variant even on the (theoretically) motion-free data, which again may be attributed to the side benefit of data and feature augmentation of the alignment branch $G_a$.
Meanwhile, the full model also slightly outperforms the no\_smooth variant.
Last but not least, {\color{blue}our hybrid 2D-3D architecture significantly outperforms the 3D-3D counterpart with appreciable margins, suggesting that the former can better handle the anisotropic OCT volumes even in the absence of obvious motion artifact.}

{\color{blue}\subsubsection{Performance on data with severe pathology}
The Duke DME dataset contains patients with severe DME pathology, especially for the ones in the testing split with damaged retinal structures by large pathological regions.
Therefore, we use the DME dataset to assess the applicability of our method to the group of data where large variations can be caused by severe pathologies.
In addition, we compare the performance of our method to that of several exiting methods which were also evaluated on the dataset, including FCBR \citep{he2019fully,he2021structured} and three SOTA graph-based methods: \cite{Chiu:15}, \cite{karri2016learning}, and \cite{rathke2017locally} (results of comparing methods are from \cite{he2019fully,he2021structured} based on the same data split and evaluation protocol).
The results are shown in Table \ref{tab:dmeresult}.
{\color{purple}As we can see, our method achieves the lowest overall MAD averaged over eight surfaces, apparently outperforming FCBR and other methods by modest (with a 0.33 $\upmu$m advantage) and substantial (with 1.34--3.17 $\upmu$m advantages) margins, respectively.
It also yields the lowest MADs for five surfaces.
Notably, for the three surfaces commonly disrupted by DME (IPL-INL, INL-OPL, and OPL-ONL), e.g., in disorganization of the retinal inner layers \citep{sun2014disorganization}, our method demonstrates apparent improvements over the existing SOTA.}
These results validate our method's applicability and efficacy on OCT data with severe pathology, too.
}
{\color{purple}Fig. \ref{fig:seg_DME} shows example segmentations by our method of two B-scans in the Duke DME dataset.}

\subsection{Semi-supervised results with sparse annotation}

To simulate sparse annotation at different degrees of sparseness, we sample the original slice-wise annotation evenly with varying fraction factors.
For example, a fraction factor of 1/8 means we take only one B-scan's annotation for every eight  B-scans.
Note that to avoid unwanted boundary effect, annotations of the first and last B-scans of an OCT volume are always included.
In the extreme case, only three B-scans are annotated for an OCT volume, i.e., the first, last, and middle ones. 
{\color{blue}As it is not straightforward to optimally extend the SOTA fully supervised methods (i.e., RelayNet, MGU-Net, FCBR, IPM, and DDP) for the semi-supervised settings, we only use the annotated B-scans for their training.
In addition, we further include an uncertainty-guided semi-supervised method exclusively developed for OCT layer segmentation, namely U-SLS \citep{sedai2019uncertainty}, for comparison in the semi-supervised settings.}
We implement and empirically optimize U-SLS.

The results on the A2A and JHH datasets are shown in Fig. \ref{semi_result_a2a}.
The performance of all methods degrades with the decreasing number of annotated B-scans, as expected.
Yet the extent of degradation varies, and for all the evaluated fraction factors (from 1/4 to 1/24) our method maintains the best performance of all methods on both datasets.
{\color{yellow}Specifically, the performance of our method is relatively stable for fraction factors down to 1/12}, and its advantage over other methods becomes most prominent in the extreme case of three annotations (1/20 and 1/24 faction factors on the A2A and JHH datasets, respectively).
{\color{purple}With equal and less than 1/8 of the B-scans annotated, our method is significantly better than all other ones on both datasets (except for DDP in the 1/8 setting on the JHH dataset), as indicated by the Wilcoxon signed-rank test.}
Notably, the performance of our method with only three annotations is equal to or better than that of ReLayNet \citep{roy2017relaynet}, {\color{yellow}MGU-net} \citep{li2021multi} and U-SLS \citep{sedai2019uncertainty} with full annotations on both datasets, suggesting its practical usability with sparse annotation.
We attribute the superior performance of our method
to the effective use of unannotated B-scan images by enforcing 3D surface coherence of the retinal layers, and the coupling of B-scan layer segmentation and motion correction.

\section{Discussion and conclusion}
This work presented a novel hybrid 2D-3D framework for simultaneous B-scan alignment and retinal surface regression of volumetric OCT data, which was applicable and proved effective to both fully and semi-supervised settings.
The core idea behind our framework was the global coherence of the retinal layer surfaces both within and across-B-scan.
Experimental results on {\color{blue}three} public clinic datasets and a synthetic dataset showed that our framework could effectively align the B-scans for motion correction and that it was superior to existing state-of-the-art methods for retinal layer segmentation in both fully and semi-supervised settings.
Also, the ablative experiments verified the efficacy of the design and newly proposed modules of our framework.

{\color{purple}The core motivation of this work was that smoothness was an intrinsic property of the retinal layers.
Correspondingly, we proposed two losses to make use of the natural smoothness: the supervised B-scan alignment loss $\mathcal{L}_\mathrm{SmoothA}$ and the regulating global coherence loss $\mathcal{L}_\mathrm{SmoothS}$.
The efficacy of these two losses was validated by the ablation experiments, contributing to the superior performance of our method.
Further, these two losses also enabled our framework to use B-scans of sparsely annotated OCT volumes for effective semi-supervised segmentation.}
In contrast to FCBR \citep{he2019fully}, which was among the previous best-performing fully supervised methods and only paid attention to the intra-B-scan layer coherence, our framework comprehensively took into account the complete 3D layer coherence, both intra- and inter-B-scan.

{\color{purple}Our global smoothness loss $\mathcal{L}_\mathrm{SmoothS}$ took the form ${\sum}_{b=1}^{N_B}{\sum}_{a=1}^{N_A}\big\|\triangledown \hat{S}(b,a)\big\|^2$, which can be dated to the classical Mumford-Shah functional \citep{mumford1989optimal}.
When functioning alone, it preferred a flat surface $\hat{S}=c$, where $c$ is a constant.
In practice, such loss is almost always used with other loss function(s) as a regulating term, e.g., our overall optimization objective in Eqn. (\ref{eq:L_seg}).
Appropriately weighed in this case, $\mathcal{L}_\mathrm{SmoothS}$ encouraged locally constant and slowly varying surfaces as a compromise, which was also our desirable notion of ``smoothness'' in this work. 
Given this notion, most foveal pit and pathology regions can be considered smooth for their slowly happening transitions and relative local constancy. 
Note that this notion of smoothness could also handle sudden jumps between normal tissue and severe pathology (i.e., edges), where piece-wise smooth surfaces on both sides of the edges were preferred to a single flat surface due to the balanced effects of the various loss terms.
This was because, with proper weights, the decreases in other losses outweighed the increase in $\mathcal{L}_\mathrm{SmoothS}$ due to the jump on the edge.
Fig. \ref{fig:seg_JHH}, Fig. \ref{fig:seg_A2A}, and Fig. \ref{fig:seg_DME} showed examples of successful segmentation by our framework in (1) the foveal pit region, (2) an AMD case with drusen, and (3) {\color{purple}DME cases with minor to severe pathology regions}, respectively.
Also, the quantitative evaluation results in Table 3, Table 4, and Table 6 showed that the performance of our framework was not appreciably affected by these regions.
Therefore, our framework generally worked well in the presence of 3D incoherence/discontinuity due to the natural structure and pathology of the retinal layers.}

{\color{purple}Meanwhile, small, early disruptions in the layer boundaries that are in the scale of artifacts may exhibit differently.
On the one hand, the possibilities were low for such disruptions to be mistaken for misalignment artifacts by our framework. 
As we only considered B-scan-wise realignment, the impact of local disruptions of a few layers could be effectively mitigated by most other layers. 
On the other hand, it was possible that such subtle disruptions might be smoothed out if the smoothness constraint was overly emphasized by an improper weight and thus harmful.
In this work, we empirically found that setting the weights of $\mathcal{L}_\mathrm{SmoothS}$ for different layer surfaces according to their extents of natural smoothness estimated from ground truth segmentation worked well for all three evaluated datasets.
For potential application to datasets of primarily more subtle, early pathology, we caution that it may be necessary to identify optimal weights with more rigorous/advanced techniques such as grid search. 
In addition, it would be interesting to explore piece-wise smoothness constraint, which explicitly accommodates edges.}

Coupling the B-scan alignment and layer segmentation also contributed to the superior performance of our method.
On the A2A dataset, which was subject to appreciable motion artifact (Table \ref{tab:abalate} top), pre-aligning the OCT volumes was 0.1 $\upmu$m short compared to the proposed framework in overall MAD.
We conjecture this was due to the side benefit of data and feature augmentation of the alignment branch $G_a$.
Meanwhile, on theoretically motion-free data (the JHH dataset; Table \ref{tab:abalate} bottom), incorporating the alignment branch did not harm the segmentation accuracy, but improved it slightly.
This is desirable as we can employ a unified framework for volumetric OCT data with and without motion artifact.
Lastly, our method also outperformed previous, dedicated motion correction methods.

\begin{table}[t]
\color{blue}
	\caption{\color{blue}Comparison of model complexity and inference time with three existing best performing methods: FCBR \citep{he2019fully,he2021structured}, IPM \citep{xie2022globally}, and DDP \citep{xie2022deep}.
    The inference time is evaluated by aligning and segmenting the OCT volumes from the A2A dataset using an NVIDIA 2080 Ti GPU.
    For fair comparison, the inference times of comparing methods include pre-alignment by NoRMCorre \citep{pnevmatikakis2017normcorre}.}
    \centering

\begin{adjustbox}{width=.95\columnwidth}

\begin{tabular}{rcccc}
\hline
                          & FCBR & IPM    & DDP    & Proposed \\ \hline
Num. parameters (million) & 1.07 & 134.49 & 134.50 & 4.28     \\
Inference time (second)        & 1.81 & 1.76   & 1.92   & 2.75     \\ \hline
\end{tabular}
\end{adjustbox}
	\label{tab:param_time}
\end{table}

{\color{blue}Considering that our hybrid 2D-3D framework included two 3D encoders---one for segmentation and the other for B-scan alignment, we investigate its model complexity and inference efficiency compared to existing best performing methods which all employed pure 2D networks: FCBR \citep{he2019fully,he2021structured}, IPM \citep{xie2022globally}, and DDP \citep{xie2022deep}.
As shown in Table \ref{tab:param_time}, our model had more parameters than FCBR (4.28  versus 1.07 millions)---as expected, but the difference (in theory about 12 MiB memory for commonly used single-precision floats) was negligible on most modern hardware.
Meanwhile, IPM and DDP employed deeper networks of significantly larger scales with more layers and feature channels, both having $\sim$134.5 million parameters.
As to the inference efficiency, our method was less than a second slower than the other methods (2.75 versus 1.81, 1.76, and 1.92 seconds).
We regard the marginally slower speed of our method as an acceptable trade-off in practice, especially to applications where 3D continuity or annotation cost is crucial.
}

This work also had some limitations.
First, like the existing approaches to B-scan motion correction which did not require any additional reference image, our motion correction could not guarantee restoration of the true retinal curvature.
However, not depending on any extra image acquisition, our method can be readily applied to existing archive data.
{\color{purple}Second, this work implemented a primitive method for transverse motion correction, although it empirically worked well on the specific data used in this study. In the future, we may need to develop more sophisticated, integrated techniques for potential data unlike those in this work.}

{\color{purple}Finally, we note from our literature search that compared with the active research on pushing the frontier of 2D layer segmentation in retinal OCT images with the significant development of deep neural networks, research on motion artifact correction has lagged behind. 
Therefore, we advocate more attention to the latter to facilitate effective 3D analysis of retinal OCT images to the community and consider this work a step forward.}

\section*{Acknowledgments}
This work was supported in part by the Ministry of Science and Technology of the People's Republic of China (2021ZD0201900, 2021ZD0201903) and in part by the Shenzhen Basic Research Program (JCYJ20190809120205578).


\end{document}